\documentclass[11pt]{article}
\usepackage{amssymb}
\usepackage{color}
\usepackage{pstricks,pst-node}
\usepackage{pst-plot}
\catcode`\@=11
\def\marginnote#1{}

\newcount\hour
\newcount\minute
\newtoks\amorpm
\hour=\time\divide\hour by60 \minute=\time{\multiply\hour by60
\global\advance\minute by-\hour}
\edef\standardtime{{\ifnum\hour<12 \global\amorpm={am}%
        \else\global\amorpm={pm}\advance\hour by-12 \fi
        \ifnum\hour=0 \hour=12 \fi
        \number\hour:\ifnum\minute<10 0\fi\number\minute\the\amorpm}}
\edef\militarytime{\number\hour:\ifnum\minute<10 0\fi\number\minute}

\def\draftlabel#1{{\@bsphack\if@filesw {\let\thepage\relax
      \xdef\@gtempa{\write\@auxout{\string
          \newlabel{#1}{{\@currentlabel}{\thepage}}}}}\@gtempa \if@nobreak
    \ifvmode\nobreak\fi\fi\fi\@esphack} \gdef\@eqnlabel{#1}}
    \def\@eqnlabel{}
\def\@vacuum{}
\def\draftmarginnote#1{\marginpar{\raggedright\scriptsize\tt#1}}

\def\draft{
  \oddsidemargin -.5truein
  \def\@oddfoot{\footnotesize \sl preliminary draft \hfil
    \rm\thepage\hfil\sl\today\quad\militarytime}
  \let\@evenfoot\@oddfoot \overfullrule 3pt
    \let\label=\draftlabel
    \let\marginnote=\draftmarginnote
  \def\@eqnnum{(\theequation)\rlap{\kern\marginparsep\tt\@eqnlabel}%
    \global\let\@eqnlabel\@vacuum}

  }

\hoffset= - 1.0in

\def\be{\begin{equation}}
\def\ee{\end{equation}}
\def\bea{\begin{eqnarray}}
\def\eea{\end{eqnarray}}
\def\<{\langle}
\def\>{\rangle}

\def\res{{{\rm res}}}

\def\d{\partial}
\def\N2{${\cal N}=2$}

\def\tr{{\mathrm{tr\,}}}

\def\1N{${\cal N}=1$}
\def\4N{${\cal N}=4$}

\def\e{{\,\rm e}\,}

\def\bea{\begin{eqnarray}}
\def\eea{\end{eqnarray}}
\def\bqa{\begin{eqnarray}}
\def\eqa{\end{eqnarray}}

\def\beq{\begin{equation}}
\def\eeq{\end{equation}}
\def\ba{\beq\begin{array}{c}}
\def\ea{\end{array}\eeq}
\def\be{\beq}
\def\ee{\eeq}

\let\text=\mathrm

\newcommand\theTag[1]{(\ref{#1})}
\def\e{e}

\def\beq{\begin{equation}}
\def\eeq{\end{equation}}
\def\bea{\begin{eqnarray}}
\def\eea{\end{eqnarray}}

\renewcommand{\d}{{{\partial}}}

\renewcommand{\<}{\langle}
\renewcommand{\>}{\rangle}

\def\2{{1\over 2}}

\def\d{\partial}

\def\â{$\tau$}

\newcommand{\cpict}[3]{
\dimen1=#1\advance\dimen1 by-\hsize\divide\dimen1 by-2 \vtop to #2{
\noindent\hskip\dimen1{\special{em:graph #3.bmp}} \vfil}\hskip-2cm }

\newcommand{\dV}{\frac{\partial}{\partial V(p)}}

\newcommand{\iint}{\int\!\!\int}

\newcommand{\tcg}{\textcolor{brown}}

\let\@@savethanks\thanks
\def\thanks#1{\gdef\thefootnote{\alph{footnote}}\@@savethanks{#1}}

\title{{\bf Logarithmic potential $\beta$-ensembles and Feynman graphs} \vspace{.5cm}}
\author{{\bf L. Chekhov}\thanks{E-mail: \ chekhov@mi.ras.ru}\date{ }
\\
{\it Steklov Mathematical Institute, ITEP}, and {\it Laboratoire Poncelet, Moscow, Russia}\\
{\it Concordia University, Montr\'eal, Canada}}

\begin{document}
\maketitle

\vspace{-7cm}

\begin{center}
\hfill ITEP/TH-33/10
\end{center}

\vspace{7cm}

\noindent
{\sl To my teacher, Andrei Alexeevich Slavnov, who showed me the beauty of Feynman graphs}

\begin{abstract}
We present the diagrammatic technique for calculating the free
energy of the matrix eigenvalue model (the model with arbitrary
power $\beta$ by the Vandermonde) to all orders of $1/N$ expansion
in the case where the limiting eigenvalue distribution spans
arbitrary (but fixed) number of disjoint intervals (curves)
and when logarithmic terms are present. This diagrammatic
technique is corrected and refined as compared to our first paper
with B.Eynard of year 2006.
\end{abstract}
\def\thefootnote{\arabic{footnote}}

\section{Introduction}
The standard Hermitian one-matrix model is determined, after integrating out angular
degrees of freedom, by the $N$-fold integral over eigenvalues of the form
$$
\int\prod_{i=1}^N dx_i\, \Delta(x)^{2}\e^{-\frac{1}{\hbar}\sum_{i=1}^N V(x_i)},\quad \hbar=t_0/N
$$
where $\Delta(x)$ is the Vandermonde determinant of eigenvalues $x_i$, $\hbar$ is the formal expansion
parameter, and $t_0$ is the normalized number of eigenvalues.
The $\beta$-ensembles, or $\beta$-eigenvalue models, are generalizations of this integral
obtained by replacing the exponent of the Vandermonde determinant by an arbitrary positive
number $2\beta$. In the proper normalization, we then evaluate the (formal) integral
\be
\int\prod_{i=1}^N dx_i\, |\Delta(x)|^{2\beta}\e^{-\frac{\sqrt{\beta}}{\hbar}\sum_{i=1}^N V(x_i)}
=\e^{-\mathcal F}.
\label{beta-pol}
\ee
The three values $\beta=1/2,1,2$ correspond in the Wigner classification
by the respective ensembles of orthogonal, Hermitian, and self-dual quaternionic matrices, but
we can put forward the problem of calculating the perturbation expansion of the $\beta$-model free energy ${\mathcal F}$
for arbitrary value of $\beta$.

In our joint paper with B.~Eynard~\cite{ChEy-b}, we developed the perturbation expansion solution to the integral (\ref{beta-pol}).
Our procedure enables constructing all the correlation function and the free energy expansion terms order by order of the
double asymptotic expansion in the two parameters: $\hbar^2$ and $(\sqrt{\beta}-\sqrt{\beta^{-1}})\hbar$. The terms of these
expansions (for all the terms except few leading terms) were expressed using the special Feynman-like diagrammatic technique,
which is a generalization of the diagrammatic technique originally constructed in our paper~\cite{ChEy} for the
Hermitian one-matrix model.

The recent interest to the $\beta$-eigenvalue models is due to the Alday, Gaiotto and Tachikawa
conjecture~\cite{AGT} relating instantonic generating functions to the conformal blocks of the Liouville theory
(see also~\cite{MMM}). These
conformal blocks were in turn associated with the asymptotic limits of the $\beta$-ensembles containing
logarithmic terms in the potential. More precisely, it was shown (see~\cite{Eguchi} and references therein)
that the Penner-like model with three logarithmic terms proposed by Dijkgraaf and Vafa~\cite{DV09}
correctly reproduces the standard results of $N=2$ supersymmetric gauge theories.

On the other hand, explicit calculations performed by Sara Pasquetti and Marcos Mari{\~n}o had revealed that applying the original
diagrammatic technique of~\cite{ChEy-b} results in a nonsymmetric result for the two-point correlation function already in the subleading
expansion order. This clearly indicated the incompleteness of the diagrammatic technique of our first paper.

Therefore, the aim of this paper is twofold: on the one hand, we extend the pattern of the paper~\cite{ChEy-b} including the logarithmic
potentials into consideration and, on the other hand, we formulate a new {\sl improved diagrammatic technique} endowed
with the new selection rules that excludes part of diagrams that caused a mismatch when calculating correlation functions.

As a nearest perspective, it is
interesting to include the hard edges into consideration, i.e., to construct the consistent diagrammatic technique for the
$\beta$-ensembles in the situation where we put explicit restrictions on the intervals of the eigenvalue distribution; the first step in
this direction was done in the recent paper~\cite{BEMN} in which a one-cut solution was investigated and results for few lower order terms of
the free-energy expansion were derived.

\section{Eigenvalue models in $1/N$-expansion \label{s:1MM}}

In this section we show how the technique of Feynman graph expansion elaborated in the
case of Hermitian one-matrix model~\cite{Ey1},~\cite{ChEy} can be adapted to solving the
(formal) eigenvalue model with the action
\be
\int\prod_{i=1}^N dx_i\, \Delta^{2\beta}\e^{-\frac{\sqrt{\beta}}{\hbar}\sum_{i=1}^N {\widehat V}(x_i)}
\prod_{k=1}^t\left(\prod_{i=1}^N|x_i-\alpha_k|^{\frac{\sqrt{\beta}}{\hbar}\gamma_k}\right)=\e^{-\cal F},
\label{X.1}
\ee
where ${\widehat V}(x)=\sum_{s\geq 1}^{}t_sx^s$, $\hbar = {t_0/ N}$ is
a formal expansion parameter, and $t_0$ is the normalized eigenvalue number.
The integration in (\ref{X.1}) goes
over $N$ variables $x_i$ having the sense of eigenvalues of the
Hermitian matrices for $\beta=1$, orthogonal matrices for $\beta=1/2$,
and symplectic matrices for $\beta=2$. In what follows, we set $\beta$ to be
arbitrary positive number. The
integration may go over curves in the complex plane of each
of $N$ variables $x_i$. For $\beta\ne1$, no topological expansion
in even powers of $\hbar$ exists and we rather have the expansion in all
integer powers of $\hbar$. We also assume
the potential ${\widehat V}(p)$ to be a polynomial
of the fixed degree $m+1$, or, if we include logarithmic terms in the potential introducing
\be
\label{totalV}
V(x_i)={\widehat V}(x_i)+\sum_{k=1}^t\gamma_k\log |x_i-\alpha_k|,
\ee
then it suffices to demand the derivative $V'(p)$ to be a rational function~\cite{Ey1}.

For brevity, we let the symbol $\tr f(X)$ to denote $\sum_{i=1}^N f(x_i)$ for any function $f(x)$.

\subsection{The loop equation and resolvents}\label{ss:loop}

The averages corresponding to partition function~\theTag{X.1} are
defined in a standard way:
\beq
\bigl\langle F(X)\bigr\rangle=
\frac1Z\int_{N}DX\,F(X)\,\exp\left(-{\sqrt{\beta}t_0\over \hbar}\tr V(X)\right)
\label{X.3*}
\eeq
with the total potential $V(x)$,
and we introduce their formal
generating functionals: the one-point resolvent
\beq
W(p)=
\hbar\sqrt{\beta}
\sum_{k=0}^{\infty}
\frac{\langle\tr X^{k}\rangle}{p^{k+1}}
\label{X.3}
\eeq
as well as the $s$-point resolvents $(s\geq2)$
\beq
W(p_1,\dots,p_s)=
(\hbar\sqrt{\beta})^{2-s}
\sum_{k_1,\dots,k_s=1}^{\infty}
\frac{\langle\tr X^{k_1}\cdots\tr X^{k_s}\rangle_{\mathrm{conn}}}
{p_1^{k_1+1}\cdots p_s^{k_s+1}}=
\hbar^{2-s}
\left\langle\tr\frac{1}{p_1-X}\cdots
\tr\frac{1}{p_s-X}\right\rangle_{\mathrm{conn}}
\label{X.4}
\eeq
where the subscript ``$\mathrm{conn}$" pertains to the connected
part.

These resolvents are obtained from the free energy ${\cal F}$ through the
action
\bea
W(p_1,\dots,p_s)&=&-\hbar^2\frac{\d}{\d V(p_s)}\frac{\partial}{\partial V(p_{s-1})}\cdots
\frac{\partial {\cal F}}{\partial V(p_1)}=\nonumber
\\
&=&\frac{\partial }{\partial V(p_s)}\frac{\partial }{\partial V(p_{s-1})}\cdots
\frac{\partial }{\partial V(p_2)}W(p_1),
\label{X.5}
\eea
of the {\em loop insertion operator}
\beq
\frac{\partial }{\partial V(p)}\equiv
-\sum_{j=1}^{\infty}\frac{1}{p^{j+1}}\frac{\d}{\d t_{j}}.
\label{X.6}
\eeq
Therefore, if one knows exactly the one-point resolvent for arbitrary
potential, all multi-point resolvents can be calculated by induction.

The loop insertion operator has an interpretation in the theory of symmetric forms.
Let us set into the correspondence to an $s$-point correlation function $W_{p_1,\dots,p_s}$
the symmetric form $W_{p_1,\dots,p_s}dp_1\cdots dp_s$ that is meromorphic on our Riemann surface. We let $\Omega^s_{1,0}$ denote
the (infinite-dimensional) space of such forms. Then, the action of the loop insertion operator increases the order
of this form by one,
\beq
\frac{\partial}{\partial V}\,: \Omega^s_{1,0}\mapsto \Omega^{s+1}_{1,0}.
\label{dV-form}
\eeq

We now consider the perturbation expansion of quantities in (\ref{X.5}).
In the above normalization, the $\hbar$-expansion has the form
\beq
W(p_1,\dots,p_s)=\sum_{r=0}^{\infty}
\hbar^{r}
W_{r/2}(p_1,\dots,p_s),\quad s\geq1,
\label{X.7}
\eeq
and for $\beta\ne 1$ it comprises both even and odd powers of $\hbar$, whereas in the
Hermitian matrix model case ($\beta=1$) the corresponding expansion (the so-called genus expansion) goes over only
even powers of $\hbar$.

We obtain the {\em master loop equation} of the eigenvalue model (\ref{X.1}) if
we perform the changing $x_i\to x_i+\frac{\epsilon}{x_i-p}$ of the integration variable.
The resulting ({\bf exact}) equation first takes the following form in terms of the means:
\bea
&&-\sum_{i,j=1}^N\left\langle\frac{\beta}{(x_i-p)(x_j-p)}\right\rangle-
\sum_{i=1}^N\left\langle\frac{1-\beta}{(x_i-p)^2}\right\rangle
+\frac{\sqrt{\beta}}{\hbar}{\widehat V}'(p)\sum_{i=1}^N\left\langle\frac{1}{x_i-p}\right\rangle
\nonumber
\\
&&+\sum_{i=1}^N\left\langle\sum_{s=1}^{m+1} st_{s}\frac{x^{s-1}_i-p^{s-1}}{x_i-p}\right\rangle
-\sum_{k=1}^t\frac{\sqrt{\beta}}{\hbar}\gamma_k\frac{1}{\alpha_k-p}\sum_{i=1}^N\left[\left\langle\frac{1}{x_i-\alpha_k}\right\rangle
-\left\langle\frac{1}{x_i-p}\right\rangle\right]=0.
\label{mean}
\eea
We let $P_{m-1}(p)$ denote the degree-$(m-1)$ polynomial
$\sum_{i=1}^N\left\langle\sum_{s=1}^{m+1} st_{s}\frac{x^{s-1}_i-p^{s-1}}{x_i-p}\right\rangle$. Note also that
the resolvent $W(p)$ must be nonsingular at $p=\alpha_k$ for any $k$. Then Eq.~(\ref{mean}) becomes in normalization (\ref{X.3}):
\bea
&&-W^2(p)-\hbar^2W(p,p)+\hbar\left(\sqrt{\beta}-\sqrt{\beta}^{-1}\right)W'(p)+{\widehat V}'(p)W(p)+P_{m-1}(p)-
\nonumber
\\
&&\qquad\qquad-\sum_{k=1}^t\gamma_k\frac{1}{\alpha_k-p}\left[W(\alpha_k)-W(p)\right]=0.
\label{X.8}
\eea

\subsection{Solution at large $N$}

Here we specially consider the limit as $\hbar\to0$ in (\ref{X.8}). Note that $W^{(0)}(p)|_{p\to\infty}=t_0/p+O(p^{-2})$.

For $W^{(0)}(p)$ we have the algebraic equation
to resolve which it is useful to introduce the function
\beq
y(p)=W(p)-\frac12\left[{\widehat V}'(p)-\sum_{k=1}^t\gamma_k\frac{1}{p-\alpha_k}\right]:=W(p)-\frac12 V'(p)
\eeq
for which Eq. (\ref{X.8}) gives the full square:
\beq
y(p)=\frac{\sqrt{P_{2m+2t}(p)}}{\prod_{k=1}^t(p-\alpha_k)}.
\eeq
The function $y(p)$ may have up to $2m+2t$ branching points thus defining a hyperelliptic Riemann surface of maximum genus $m+t-1$.
We let $2n$ be the actual number of the distinct branching points $\mu_r$ (some branching points may merge producing double points),
and the actual genus is then $n-1$. We assume that we have therefore $n$ disjoint intervals $A_i$ of eigenvalue distribution and
$n-1$ ``forbidden zones'' $B_i$. We treat $n-1$ among $n$ intervals $A_i$ as a-cycles and we then let $\sum_{j=i}^{n-1}B_j$ to be
the corresponding $b_i$-cycle (one half of which goes along the physical sheet and the other half goes along the second, unphysical
sheet). So, in general, we assume that $y(p)$ has the form
$$
y(p)=M_{m+t-n}(p)\frac{\sqrt{\prod_{r=1}^{2n}(p-\mu_r)}}{\prod_{k=1}^t(p-\alpha_k)},
$$
where the polynomial $M_{m+t-n}(p)$ arises from merging branching points of $P_{2m+2t}(p)$.

The $m+t-n+1$ coefficients of the polynomial $M_{m+t-n}(p)$ and the $2n$ branching points $\mu_r$
are to be determined from the following conditions:
\begin{itemize}
\item (asymptotic condition at infinity)
$$
\lim_{p\to\infty}y(p)=-\frac12 {\widehat V}'(p)+\left(\frac12\sum_{k=1}^t \gamma_k+t_0\right)\frac1p+O(p^{-2})\qquad \hbox{$m+2$ conditions};
$$
\item (regularity at $p=\alpha_k$ of $W(p)$):
$$
\frac{\sqrt{P_{2m+2t}(\alpha_k)}}{\prod_{l\ne k}(\alpha_k-\alpha_l)}=\frac12 \gamma_k\qquad \hbox{$t$ conditions};
$$
\item (equality of chemical potentials on all cuts)
$$
\oint_{B_i}ydp=0\qquad \hbox{$n-1$ conditions}.
$$
\end{itemize}

Note that, in the context of the Seiberg--Witten theory, the last $n-1$ conditions can acquire other forms,
for instance,
\begin{itemize}
\item (fixing filling fractions on the intervals of eigenvalue distribution)
$$
\oint_{A_i}ydp=\epsilon_i\qquad \hbox{$n-1$ conditions}.
$$
\end{itemize}

So, in total we have exactly the desired number of $m+t+n+1$ conditions.

The recurrent procedure in the subsequent sections is such that all the geometry properties
of the theory (Bergmann kernels, etc.)
is completely determined by the {\em reduced} hyperelliptic Riemann surface
$$
\tilde y^2=\prod_{r=1}^{2n}(p-\mu_r),
$$
where we lose all the information on the rational function $Q(p):=M_{m+t-n}(p)/\bigl(\prod_{k=1}^t(p-\alpha_k)\bigr)$.

{\sl All the rest of the text is totally insensitive to the nature of the rational function $Q(p)$.}

We need to introduce the {\sl Bergman kernel}: the symmetric ($B(p,q)=B(q,p)$)
bi-differential that has the form
$$
B(p,q):=B(x(p),x(q))dx(p)dx(q)\sim \frac{dx(p)dx(q)}{(x(p)-x(q))^2}\hbox{ as $x(p)\to x(q)$, }\oint_{B_j}B(\cdot,p)=0
$$
in the local coordinate $x(p)$.

We also need its primitive (a one-differential in $r$):
$$
dE_{p,q}(r):=\int_{x(p)}^{x(q)}B(\cdot,r).
$$

Note that both $B(p,q)$ and $dE_{p,q}(r)$ depend only on ``reduced'' Riemann surface (are determined completely by
the branching points $\mu_r$). Definitely, they take the simplest form for {\bf one-cut} solutions: if we have just
two branching points $\mu_1$ and $\mu_2$, then, for $\bar q$ being the image of $q$ on the other sheet,
$$
dE_{q,\bar q}(r)=\frac{\sqrt{(q-\mu_1)(q-\mu_2)}}{(r-q)\sqrt{(r-\mu_1)(r-\mu_2)}}dr;
$$
for higher-genus spectral curves, this expression contains also projections terms proportional to Abelian differentials
normalized to $B$-cycles.

\subsection{Higher-order corrections}

The $\beta$-dependence enters (\ref{X.8}) only through the combination
\be
\zeta=\sqrt{\beta}-\sqrt{\beta^{-1}},
\label{X.8*}
\ee
and, assuming $\beta\sim O(1)$, we have the free energy expansion
of the form
\be
{\cal F}\equiv {\cal F}(\hbar, \zeta, t_0, t_1, t_2, \dots)
=\sum_{k=0}^{\infty}\sum_{l=0}^{\infty}{\hbar}^{2k+l-2}\zeta^l{\cal F}_{k,l}.
\label{X.2}
\ee

Substituting expansion~\theTag{X.7} in Eq.~\theTag{X.8}, we find
that $W_g(p)$ for $g\geq1/2$ satisfy the equation
\bea
-2y(p)W_g(p)&=&\sum_{g'=1/2}^{g-1/2}
W_{g'}(p)W_{g-g'}(p)+\frac{\partial }{\partial V(p)}W_{g-1}(p)+\zeta \frac{\d}{\d p}W_{g-1/2}(p)
\nonumber
\\
&&+P^{(g)}_{m-2}(p)+\sum_{k=1}^t\gamma_k\frac{1}{p-\alpha_k} W_{g}(\alpha_k)
\label{X.9}
\eea
In Eq.~\theTag{X.9}, $W_g(p)$ is expressed through only the
$W_{g_i}(p)$ for which $g_i<g$. This fact permits
developing the iterative procedure.

In analogy with (\ref{X.2}), it is convenient to expand multiresolvents $W_g(\cdot)$ in $\zeta$:
\be
W_g(p_1,\dots,p_s)=\left\{
\begin{array}{ll}
  \sum_{l=0}^{g}\zeta^{2l}W_{g-l,\ 2l}(p_1,\dots,p_s), & g\in{\mathbb Z} \\
  \sum_{l=0}^{g-1/2}\zeta^{2l+1}W_{g-l-1/2,\ 2l+1}(p_1,\dots,p_s), & g\in{\mathbb Z}+1/2 \\
\end{array}
\right.
\label{X.11}
\ee
Then, obviously, (\ref{X.9}) becomes
\bea
-2y(p)W_{k,l}(p)&=&\sum_{k_1\ge0,l_1\ge0\atop k_1+l_1>0}
W_{k_1,l_1}(p)W_{k-k_1,l-l_1}(p)+\frac{\partial }{\partial V(p)}W_{k-1,l}(p)+\zeta \frac{\d}{\d p}W_{k,l-1}(p)
\nonumber
\\
&&+P^{(k,l)}_{m-2}(p)+\sum_{s=1}^t\gamma_s\frac{1}{p-\alpha_s} W_{k,l}(\alpha_s).
\label{X.12}
\eea

Recall that because
\be
W_{k,l}(p)|_{p\to\infty} = \frac{t_0}{p}\delta_{k,0}\delta_{l,0}+O({1}/{p^2}),
\label{Winf}
\ee
all $W_{k,l}(p)$ are total derivatives,
\be
\label{total}
W_{k,l}(p)=\dV {\cal F}_{k,l},\quad k,l\ge 0.
\ee

The planar limit solution for $W_{0,0}(p)$ exactly coincides with the one in the one-matrix model.
The normalizing conditions seem to be those ensuring the coincidence of chemical potentials in all orders
of the perturbation theory
\beq\label{vanishBcycle}
\oint_{B_i} W_{k,l}(\xi,p_2,\dots,p_s) = 0,\ \ \hbox{for} \ \ k+l>0.
\eeq
Note that in the Seiberg--Witten theory, these conditions must be merely replaced by the vanishing conditions
for the $A$-cycle integrals.

The formal solution to (\ref{X.12}) is given by the integral~\cite{ChEy-b}:
\bea
W_{k,l}(p)&=&\oint_{{\mathcal C}_D^{(q)}}dq\,dE_{q,\bar q}(p)\frac{1}{2y(q)}\Bigl(
\sum_{0\le k_1\le k, 0\le l_1\le l\atop 0<k_1+l_1<k+l}
W_{k_1,l_1}(p)W_{k-k_1,l-l_1}(p)\Bigr.
\nonumber
\\
&&+\Bigl.\frac{\partial }{\partial V(p)}W_{k-1,l}(p)+ \frac{\d}{\d p}W_{k,l-1}(p)\Bigr).
\label{recursion}
\eea
Here the contour ${\mathcal C}_D^{(q)}$ lies on the physical leaf, encircles all cuts of the function $y(q)$, and leaves
outside all the other possible singularities (poles and zeros).
In the next section we develop the diagrammatic representation for terms of these recurrent relations.

\section{Graphical representation. Spatial derivative.}

The diagrammatic technique of~\cite{ChEy-b} was based on the diagrammatic technique developed in~\cite{ChEy}
for the Hermitian one-matrix model, but contained new elements. The original technique of~\cite{ChEy} contained
two types of the propagators: arrowed lines corresponding to $dE_{q,\bar q}(p)$ and nonarrowed lines corresponding
to $B(p,q)$. Simultaneously, the two-point correlation function
$$
\frac{\partial }{\partial V(p)}W_{0,0}(q)=B(p,q)-\frac{1}{(p-q)^2},
$$
and the action of the loop insertion operator on it produces the three-point correlation function described by the Rauch
variation formulas that result in the appearance of the three-point vertex denoted by light circle and containing the
integration over the contour ${\mathcal C}_D^{(q)}$ as in (\ref{recursion}) with the weight $1/(2y(q))$
First of all, the recursion kernel in (\ref{recursion}) coincides with the recursion kernel in~\cite{ChEy}, but to comply with
the new term, $\zeta \frac{\d}{\d p}W_{k,l-1}(p)$, appearing in the right-hand side, we
must introduce a {\em new} propagator $dpdy(q)$ denoted by the dashed line.

The graphical representation
for a solution of loop equation (\ref{X.9}) then looks as follows.
The multi-point resolvent $W_{k,l}(p_1,\dots,p_s)$ is represented by the block with $s$ external legs and with the
double index $k,l$.
Among these legs one, say $p_1$, is selected to be the root of maximal tree subgraph composed from arrowed propagators
and establishing a partial ordering on the set of vertices of a graph.

We present the derivative $\frac{\d}{\d p_1}W_{k,l-1}(p_1,\dots,p_s)$ as the block with $s+1$ external legs, one of which
is the dashed leg that starts also at the vertex $p_1$. For instance, for the one-point resolvent expansion term $W_{k,l}(p)$,
we obtain~\cite{ChEy-b}:
\be
{\psset{unit=.8}
\begin{pspicture}(-8,-2)(8,2)
\rput(-8,0.3){\makebox(0,0)[cc]{$p$}}
\pcline[linewidth=2pt,linecolor=brown](-8,0)(-7.5,0)
\pscircle[linewidth=1pt,fillstyle=solid,fillcolor=white](-7,0){0.5}
\rput(-7,0){\makebox(0,0)[cc]{$k,l$}}
\rput(-6,0){\makebox(0,0)[lc]{$\displaystyle=\sum\limits_{k_1{+}k_2{=}k \atop l_1{+}l_2{=}l}^{(k_i,l_i){\ne}(0,0)}$}}
\rput(-3.5,0){
\rput(0,0.3){\makebox(0,0)[cc]{$p$}}
\pcline[linewidth=2pt,linecolor=brown]{->}(0,0)(0.8,0)
\pcline[linewidth=2pt,linecolor=brown]{->}(1,0)(1.7,0.7)
\pcline[linewidth=2pt,linecolor=brown]{->}(1,0)(1.7,-0.7)
\pscircle[linewidth=1pt,fillstyle=solid,fillcolor=white](1,0){0.2}
\pscircle[linewidth=1pt,fillstyle=solid,fillcolor=white](2.2,1.2){0.7}
\pscircle[linewidth=1pt,fillstyle=solid,fillcolor=white](2.2,-1.2){0.7}
\rput(2.2,1.2){\makebox(0,0)[cc]{$k_1,l_1$}}
\rput(2.2,-1.2){\makebox(0,0)[cc]{$k_2,l_2$}}
\rput(0.8,.4){$q$}
}
\rput(0,0){
\rput(-.5,0){\makebox(0,0)[cc]{$+$}}
\rput(0,0.3){\makebox(0,0)[cc]{$p$}}
\pcline[linewidth=2pt,linecolor=brown]{->}(0,0)(0.8,0)
\psarc[linecolor=blue, linewidth=2pt](3,-2){2.82}{115}{135}
\psarc[linecolor=brown, linewidth=2pt]{->}(3,2){2.82}{225}{245}
\pscircle[linewidth=1pt,fillstyle=solid,fillcolor=white](1,0){0.2}
\pscircle[linewidth=1pt,fillstyle=solid,fillcolor=white](2.6,0){0.95}
\rput(0.8,.4){$q$}
\rput(2.6,0){\makebox(0,0)[cc]{$k{-}1,l$}}
}
\rput(4.5,0){
\rput(-.5,0){\makebox(0,0)[cc]{$+$}}
\rput(0,0.3){\makebox(0,0)[cc]{$p$}}
\pcline[linewidth=2pt,linecolor=brown]{->}(0,0)(0.8,0)
\psarc[linecolor=red,  linestyle=dashed, linewidth=2pt]{<-}(3,-2){2.82}{115}{135}
\psarc[linecolor=brown, linewidth=2pt]{->}(3,2){2.82}{225}{245}
\pscircle[linewidth=1pt,fillstyle=solid,fillcolor=white](1,0){0.2}
\pscircle[linewidth=1pt,fillstyle=solid,fillcolor=white](2.6,0){0.95}
\rput(0.8,.4){$q$}
\rput(2.6,0){\makebox(0,0)[cc]{$k,l{-}1$}}
}
\end{pspicture}
}
\label{Bertrand1}
\ee
In the case of multipoint resolvent $W_{k,l}(p_1,\dots,p_s)$ the picture is similar, we replace $p$ by $p_1$ and
add $s-1$ remaining external legs to the diagrams in the right-hand side; then, in the first diagram we must distribute the
added $s-1$ external legs over two subdiagrams in an arbitrary way and take a sum over all possible insertions that leave
the subdiagrams $W_{k_i,l_i}(q,p_{j_1},\dots,p_{j_{s_i}})$ ``stable.'' Recall that $W_{k,l}(p_1,\dots,p_s)$ is called
{\em stable} (for a nonempty set of $p_1,\dots,p_s$) if either $k>0$ or $l>0$ or $s>2$.

We can then in turn present the term $W_{k,l-1}(q,p_2,\dots,p_s)$ in the form (\ref{recursion})
with the recursion kernel $dE_{\eta,\bar \eta}(q)/(2y(\eta))$ integrated with some function $F(\eta)$, the rest of the
diagram, which we do not specify here.
Let us consider the action of the spatial derivative $\d/\d q$ on $W_{k,l-1}(q,p_2,\dots,p_s)$.
We place the {\em starting point} of the dashed directed edge at the point $q$ and associate
just $dx$ with this starting point.
The first object (on which the derivative actually acts) is
$dE_{\eta,\bar\eta}(q)$, then comes the vertex with the integration
$\frac{1}{2\pi i}\oint_{{\cal C}_{\cal D}^{(\eta)}}\frac{d\eta}{y(\eta)}$,
then the function $F(\eta)$. We can present the action of the derivative
via the contour integral around $q$ with the kernel $B(q,\xi)$:
\be
{\d\over \d {q}}\left(\oint_{{\cal C}_{\cal D}^{(\eta)}}\frac{dE_{\eta,\bar\eta}(q) d\eta}{2\pi i \ y(\eta)}F(\eta)\right)
= \mathop{{\rm Res}}_{\xi\to q} \oint_{{\cal C}_{\cal D}^{(\eta)}} \,\frac{B(q,\xi)dE_{\eta,\bar\eta}(\xi)}{d\xi}
\,\frac{d\eta}{2\pi i \ y(\eta)}\,F(\eta),
\label{Y.3}
\ee
where $q$ lies outside the integration contour for $\eta$. The integral over $\xi$ is nonsingular at infinity, so we can
deform the integration contour from ${\cal C}_{q}$ to ${\cal C}_{\cal D}^{(\xi)}>{\cal C}_{\cal D}^{(\eta)}$,\footnote{Here and hereafter,
the comparison of contours indicates their ordering.}
\beq
 \oint_{{\cal C}_{q}}\frac{B(q,\xi)dE_{\eta,\bar \eta}(\xi)}{2\pi i \ d\xi} \oint_{{\cal C}_{\cal D}^{(\eta)}}\frac{d\eta}{2\pi i \ y(\eta)}F(\eta)
= - \oint_{{\cal C}_{\cal D}^{(\xi)} >{\cal C}_{\cal D}^{(\eta)}}\frac{B(p_1,\xi)dE_{\eta,\bar \eta}(\xi)}{2\pi i \ d\xi}
\frac{d\eta}{2\pi i \ y(\eta)}F(\eta)
\label{nunu}
\eeq
We now push the contour for $\xi$ through the contour for $\eta$, picking residues at the poles in $\xi$, at $\xi=\eta$, $\xi=\bar\eta$
and at the branch points. We then obtain
\beq
\label{Y.4-1}
 - \oint_{{\cal C}_{\cal D}^{(\eta)}} \sum_\alpha \mathop{{\rm Res}}_{\xi\to \mu_\alpha}
\frac{B(q,\xi)dE_{\eta,\bar \eta}(\xi)}{ d\xi} \frac{d\eta}{2\pi i \ y(\eta)}F(\eta)
 - \oint_{{\cal C}_{\cal D}^{(\eta)}}\frac{B(q,\eta)}{2\pi i \ y(\eta)}F(\eta)
 + \oint_{{\cal C}_{\cal D}^{(\eta)}}\frac{B(q,\bar\eta)}{2\pi i \ y(\eta)}F(\eta),
\eeq
where the first integral has only
simple poles at the branch points. The main trick stems from that the residue remains unchanged if we apply the l'H\^opital rule and
replace $B(q,\xi)/dy(\xi)$ by $dE_{\xi,\bar\xi}(q)/2y(\xi)$, thus replacing $B(q,\xi)$ by
$dE_{\xi,\bar\xi}(q)dy(\xi)/2y(\xi)$\footnote{We therefore ``imitate'' pole terms by inserting $dy/y$.};
we then obtain in the right-hand side of (\ref{nunu}):
$$
 - \oint_{{\cal C}_{\cal D}^{(\eta)}} \sum_\alpha \mathop{{\rm Res}}_{\xi\to \mu_\alpha}
\frac{dE_{\xi,\bar\xi}(p_1)\,dy(\xi)\,dE_{\eta,\bar \eta}(\xi)}{ 2 y(\xi)\,d\xi}
\frac{d\eta}{2\pi i \ y(\eta)}F(\eta)
- \oint_{{\cal C}_{\cal D}^{(\eta)}}\frac{B(q,\eta)-B(q,\bar \eta)}{2\pi i \ y(\eta)}F(\eta).
$$
Pushing contour of integration for $\xi$  around the branch points back through the contour for $\eta$,
we pick residues at $\xi=\eta$ and $\xi=\bar\eta$, which both give the same contribution, and obtain
\bea
&& - \oint_{{\cal C}_{\cal D}^{(\xi)}>{\cal C}_{\cal D}^{(\eta)}} \frac{dE_{\xi,\bar\xi}(q)\,dy(\xi)\,dE_{\eta,\bar \eta}(\xi)}
{ 2\pi i \ 2 y(\xi)\,d\xi}
\frac{d\eta}{2\pi i \ y(\eta)}F(\eta) \cr
&& +  \oint_{{\cal C}_{\cal D}^{(\eta)}}  \frac{dE_{\eta,\bar\eta}(q)\,dy(\eta)}{  y(\eta)\,d\eta} \frac{d\eta}{2\pi i \ y(\eta)}F(\eta)
 - \oint_{{\cal C}_{\cal D}^{(\eta)}}\frac{B(q,\eta)-B(q,\bar\eta)}{2\pi i \ y(\eta)}F(\eta).
\eea
We evaluate the last two terms by parts therefore obtaining
\beq
 - \oint_{{\cal C}_{\cal D}^{(\xi)}>{\cal C}_{\cal D}^{(\eta)}}
 \frac{dE_{\xi,\bar\xi}(p_1)\,dy(\xi)\,dE_{\eta,\bar \eta}(\xi)}{ 2\pi i \ 2 y(\xi)\,d\xi}
 \frac{d\eta}{2\pi i \ y(\eta)}F(\eta)
 +  \oint_{{\cal C}_{\cal D}^{(\eta)}}  \frac{dE_{\eta,\bar\eta}(q)\,d\eta}{2\pi i \ y(\eta)} \,\frac{\d F(\eta)}{\d\eta}.
\eeq

We have thus found that
\bea
&& {\d\over \d_{q}}\left(\oint_{{\cal C}_{\cal D}^{(\eta)}}\frac{dE_{\eta,\bar\eta}(q) d\eta}{2\pi i \ y(\eta)}F(\eta)\right) \cr
&=& - \oint_{{\cal C}_{\cal D}^{(\xi)}>{\cal C}_{\cal D}^{(\eta)}}
\frac{dE_{\xi,\bar\xi}(q)\,y'(\xi)\,dE_{\eta,\bar \eta}(\xi)}{ 2\pi i \ 2 y(\xi)}
\frac{d\eta}{2\pi i \ y(\eta)}F(\eta)
 +  \oint_{{\cal C}_{\cal D}^{(\eta)}}  \frac{dE_{\eta,\bar\eta}(p_1)\,d\eta}{2\pi i \ y(\eta)} \,\frac{\d F(\eta)}{\d\eta},
\eea
and we can graphically present the action of the derivative as follows:
\be
{\psset{unit=.8}
\begin{pspicture}(-6,-1)(6,1)
\rput(-6,0){
\rput(-.4,0){\makebox(0,0)[cc]{$\d_{q}$}}
\pcline[linewidth=2pt,linecolor=brown]{->}(0,0)(.8,0)
\pscircle[linewidth=1pt,fillstyle=solid,fillcolor=white](1,0){.2}
\rput(0,-.4){\makebox(0,0)[cc]{$q$}}
\rput(1,-.5){\makebox(0,0)[cc]{$\eta$}}
\rput(1.2,0){\makebox(0,0)[lc]{$\Bigl\{F(\eta)\Bigr\}$}}
}
\rput(-2.5,0){\makebox(0,0)[cc]{$=$}}
\rput(-2,0){
\rput(1.4,0.8){\makebox(0,0)[cc]{$y'(\xi)$}}
\pcline[linewidth=2pt,linecolor=brown]{->}(0,0)(.8,0)
\psbezier[linewidth=2pt,linecolor=red,linestyle=dashed]{->}(0,0.8)(0.8,.7)(.8,.7)(.95,0.1)
\pcline[linewidth=2pt,linecolor=brown]{->}(1.1,0)(1.8,0)
\pscircle[linewidth=1pt,fillstyle=solid,fillcolor=black](1,0){.15}
\pscircle[linewidth=1pt,fillstyle=solid,fillcolor=white](2,0){.2}
\rput(0,-.4){\makebox(0,0)[cc]{$q$}}
\rput(1,-.5){\makebox(0,0)[cc]{$\xi$}}
\rput(2,-.5){\makebox(0,0)[cc]{$\eta$}}
\rput(2.2,0){\makebox(0,0)[lc]{$\Bigl\{F(\eta)\Bigr\}$}}
}
\rput(2.5,0){\makebox(0,0)[cc]{$+$}}
\rput(3,0){
\pcline[linewidth=2pt,linecolor=brown]{->}(0,0)(.8,0)
\pscircle[linewidth=1pt,fillstyle=solid,fillcolor=white](1,0){.2}
\rput(0,-.4){\makebox(0,0)[cc]{$q$}}
\rput(1,-.5){\makebox(0,0)[cc]{$\eta$}}
\rput(1.2,0){\makebox(0,0)[lc]{$\Bigl\{\frac{\d}{\d \eta}F(\eta)\Bigr\}$}}
}
\end{pspicture}
}
\label{Y.5-1}
\ee
In these relations and in what follows, we use light circles (``white'' vertices) to denote the integrations with
the weight $1/(2y(\eta))$ and ``black'' vertices to indicate integrations with weights containing the
first or higher derivatives $y^{(s)}(\eta)$ in the numerator.

We also assume the ordering from left to right in all the diagrams below.

We therefore see that using relation (\ref{Y.5-1}) we can push the differentiation along the arrowed
propagators of a graph. It remains to determine the action of the derivative on internal nonarrowed propagators.
But for these propagators (since it has two ends), having the term with derivative from the one side, we necessarily
come also to the term with the derivative from the other side {\em if we act by the dashed line that was
commenced before the outer end of this propagator}. Combining these terms, we obtain
$$
\d_p B(p,q)+\d_q B(p,q)=\oint_{{\cal C}_p\cup{\cal C}_q}\frac{B(p,\xi)B(\xi,q)}{2\pi i\ d\xi},
$$
and we can deform this contour to the sum of contours only around the branching points (sum of residues). Then
we can again introduce $y'(\xi)d\xi/y(\xi)$ and integrate out one of the Bergman kernels (the one that is adjacent
to the point $q$ if $p>q$ or $p$ if $q>p$; recall that, by condition, the points $p$ and $q$ must be comparable).

That is, we have
\be
{\psset{unit=.8}
\begin{pspicture}(-6,-1)(6,2)
\newcommand{\PATTERN}[1]{%
{\psset{unit=#1}
\pscircle[linewidth=1pt,fillstyle=solid,fillcolor=white](0,0){.2}
\rput(0,-.7){$r$}
\pcline[linewidth=2pt,linecolor=brown]{->}(0.2,0)(.8,0)
\rput(0.9,0){\textcolor{brown}{\hbox{\Large$\cdot$}}}
\rput(1,0){\textcolor{brown}{\hbox{\Large$\cdot$}}}
\rput(1.1,0){\textcolor{brown}{\hbox{\Large$\cdot$}}}
\pcline[linewidth=2pt,linecolor=brown]{->}(1.2,0)(1.8,0)
\pscircle[linewidth=1pt,fillstyle=solid,fillcolor=white](2,0){.2}
\rput(2,-.7){$p$}
\pcline[linewidth=2pt,linecolor=brown]{->}(2.2,0)(2.8,0)
\rput(2.9,0){\textcolor{brown}{\hbox{\Large$\cdot$}}}
\rput(3,0){\textcolor{brown}{\hbox{\Large$\cdot$}}}
\rput(3.1,0){\textcolor{brown}{\hbox{\Large$\cdot$}}}
\pcline[linewidth=2pt,linecolor=brown]{->}(3.2,0)(3.8,0)
\pscircle[linewidth=1pt,fillstyle=solid,fillcolor=white](4,0){.2}
\rput(4,-.7){$q$}
}
}
\rput(-6,0){\PATTERN{1}}
\rput(-3,0){\psarc[linecolor=blue, linewidth=2pt](0,-.58){1.16}{35}{145}}
\rput(-3,0){\psarc[linecolor=red, linestyle=dashed, linewidth=2pt]{<-}(0,-1.82){3.64}{95}{148}}
\rput(-4.5,0.5){$\partial_p{+}\partial_q$}
\rput(-1,0){$=$}
\rput(0,0){\PATTERN{1}}
\pcline[linewidth=2pt,linecolor=brown]{->}(4.2,0)(5.8,0)
\pscircle[linewidth=1pt,fillstyle=solid,fillcolor=black](6,0){.2}
\rput(6,-.7){$\eta$}
\rput(6.5,.2){$y'$}
\rput(4,0){\psarc[linecolor=blue, linewidth=2pt](0,-2){2.83}{48}{132}}
\rput(3,0){\psarc[linecolor=red, linestyle=dashed, linewidth=2pt]{<-}(0,-1.82){3.64}{32}{148}}
\end{pspicture}
}
\label{variation-1}
\ee

We must therefore improve the diagrammatic technique of $\beta$-model in comparison with
the Hermitian one-matrix model by including the dashed lines. They are not actual propagators, but
these lines ensure the proper combinatorics of diagrams.
Indeed, from (\ref{Y.5-1}) it follows that the derivative action on the ``beginning'' of the dashed line
is null, $\d_p dp=0$, and when this derivative acts on the ``end'' of this line with the black vertex, we merely have
\be
\d_q y'(q)=y''(q),
\label{Y.6-1}
\ee
which we denote symbolically as {\em two} dashed propagators ending at the same black vertex. If we continue to act
by derivatives $\d/\d q$, then, obviously,
when $k$ dashed propagators are terminated at the {\em same} vertex, we have the $k$th order
derivative $y^{(k)}(q)$ corresponding to them.

Another lesson from studying the action of the spatial derivative is that {\em not all} diagrams are possible: for example,
the diagram containing the vertex
$$
{\psset{unit=.8}
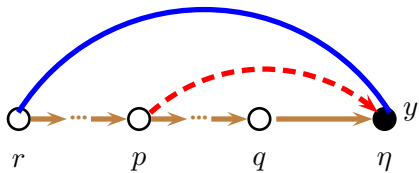
\begin{pspicture}(-4,-1)(4,2)
\newcommand{\PATTERN}[1]{%
{\psset{unit=#1}
\pscircle[linewidth=1pt,fillstyle=solid,fillcolor=white](0,0){.2}
\rput(0,-.7){$r$}
\pcline[linewidth=2pt,linecolor=brown]{->}(0.2,0)(.8,0)
\rput(0.9,0){\textcolor{brown}{\hbox{\Large$\cdot$}}}
\rput(1,0){\textcolor{brown}{\hbox{\Large$\cdot$}}}
\rput(1.1,0){\textcolor{brown}{\hbox{\Large$\cdot$}}}
\pcline[linewidth=2pt,linecolor=brown]{->}(1.2,0)(1.8,0)
\pscircle[linewidth=1pt,fillstyle=solid,fillcolor=white](2,0){.2}
\rput(2,-.7){$p$}
\pcline[linewidth=2pt,linecolor=brown]{->}(2.2,0)(2.8,0)
\rput(2.9,0){\textcolor{brown}{\hbox{\Large$\cdot$}}}
\rput(3,0){\textcolor{brown}{\hbox{\Large$\cdot$}}}
\rput(3.1,0){\textcolor{brown}{\hbox{\Large$\cdot$}}}
\pcline[linewidth=2pt,linecolor=brown]{->}(3.2,0)(3.8,0)
\pscircle[linewidth=1pt,fillstyle=solid,fillcolor=white](4,0){.2}
\rput(4,-.7){$q$}
}
}
\rput(-3,0){\PATTERN{1}}
\pcline[linewidth=2pt,linecolor=brown]{->}(1.2,0)(2.8,0)
\pscircle[linewidth=1pt,fillstyle=solid,fillcolor=black](3,0){.2}
\rput(3,-.7){$\eta$}
\rput(3.5,.2){$y'$}
\rput(0,0){\psarc[linecolor=blue, linewidth=2pt](0,-1.82){3.64}{32}{148}}
\rput(1,0){\psarc[linecolor=red, linestyle=dashed, linewidth=2pt]{<-}(0,-2){2.83}{48}{132}}
\end{pspicture}
}
$$
{\em never} appear in our approach. Later on we formulate all the selection rules for the diagrams as well as present all
types of possible vertices.

We also introduce the black-and-white coloring of vertices in accordance with the rule:
if there is no factors $y^{(k)}(q)$ standing by the vertex, it is white; if there are such factors, the vertex is painted black.

We need now to produce the action of the loop insertion operator in the new setting.
We calculate the action of $\d/\d V(r)$ on the dashed propagator. Obviously,
\be
\label{dydV}
\frac{\d}{\d V(r)}y^{(k)}(q)=\frac{\d^k}{\d q^k}B(r,q),
\ee
and any attempt to simplify this expression or to reduce it to a combination of
previously introduced diagrammatic elements fails. This means that we must consider it a {\em new} element
of the diagrammatic technique. We indicate it by preserving
$k$ dashed arrows still terminating at the vertex $q$ with the added nonarrowed solid line
corresponding to the new propagator $B(r,q)$.
The vertex then change the coloring from black to white because it contains $y^{(k)}(q)$ factors no more
and we therefore assume that these derivatives act on $B(r,q)$.

An accurate analysis demonstrates nevertheless that we must include also vertices with the mixed action of
spatial derivatives: on $B(r,q)$ and on $y^{(s)}(q)$ standing at this vertex (which is then also painted black).
These vertices were also missed in the original paper~\cite{ChEy-b}.

We leave the detailed proof for subsequent publications and present in the next section the {\em complete set of rules}
for constructing diagrammatic representation for $W_{k,l}(p_1,\dots,p_s$ for all $k$ and $l$.

\section{Feynman diagrammatic rules}\label{ss:rules}

We now describe the diagrammatic technique that results from the action of $\partial_{q}$---the spatial derivative and
the action of the loop insertion operator $\d/\d V(q)$ with describing new vertices and selection rules.

In this diagrammatic technique, all the solid arrowed propagators $dE_{q,\bar q}(p)$ are free of spatial derivatives, vertices
always contain the factor $1/(2y(q))$ and may or may not contain additional derivatives $y^{(s)}(q)$ in the numerator and the
solid nonarrowed propagators corresponding to $B$-kernels may contain additional spatial derivatives acting only on the
innermost end of the propagator (recall that $B$-propagators can join {\em only} vertices comparable in the sense of partial ordering
established by the tree of arrowed propagators.

In the graphical representation below the ordering of vertices is implied from left to right.

The diagrammatic technique for constructing resolvents $W_{k,l}(p_1,\dots,p_s)$ is as follows.

We take the sum over connected graphs containing $k$ internal nonarrowed propagators (Bergmann kernels),
$l$ dashed arrowed propagators (those are always internal ``derivatives'') and $s$ external legs such that
\begin{itemize}
\item we begin constructing each graph by constructing the maximum rooted directed tree subgraph composed from the
propagators $dE_{q,\bar q}(p)$ ($q$ here is the inner vertex with respect to $p$, the arrow is pointed toward $q$)
and take the sum over all choices of such rooted tree subgraphs;
\item we choose the external vertex $p_1$ to be the root of the tree, one and the same $p_1$ for all the graphs;
\item all other external legs are nonarrowed propagators $B$; we call propagators $dE$ and $B$ solid as we set solid lines
corresponding to them; all the external points of the graph are considered outermost points;
\item other internal edges (i.e., edges that are incident only to the graph vertices and not to $p_1,\dots,p_s$)
are either $B$ (there are exactly $k$ of them) or dashed arrowed edges (there are exactly $l$ of them);
\item rooted tree subgraph establishes a partial ordering on the set of internal vertices; {\sl we allow internal $B$ edges and
dashed edges to connect only comparable vertices}. Dashed arrows are directed always inward, but can begin and terminate at the
same vertex; if such a vertex exist, it must be an innermost vertex of the graph. The $B$-lines can also begin and terminate at the
same vertex; in this case, this vertex is also necessarily an innermost vertex of the graph.
\item A partial ordering establishes the order of taking residues: we start from innermost vertices and continues toward
the root of the tree.
\item All $s$ external legs $p_j$ are outside all the integrations.
\item All the internal vertices are incident to exactly one incoming arrowed line and can be incident up to two other ends of solid lines.
\item The vertices of the graph are painted either black or white
(depending on whether they respectively contain or do not contain the factors $y^{(k)}(q)$) and must satisfy the {\em selection rules} below.
\item All the graphs satisfying all the above properties enter the final answer with the standard symmetry coefficients that are
the reciprocal volumes of the discrete automorphism groups of the graphs.
\end{itemize}

The selection rules follow from the complete {\em list of ten types of vertices} below. Their origin will be partly clarified in the next
section when considering the action of the loop insertion operator on the elements of the diagrammatic technique. The general proof will be
published elsewhere.

{\em The vertices with three adjacent solid lines}
\be
{\psset{unit=.8}
\begin{pspicture}(-8,-1)(8,2.5)
\newcommand{\PATTERN}[1]{%
{\psset{unit=#1}
\pscircle[linewidth=1pt,fillstyle=solid,fillcolor=white](0,0){.2}
\pcline[linewidth=2pt,linecolor=brown]{->}(0.2,0)(.7,0)
\rput(0.8,0){\textcolor{brown}{\hbox{\Large$\cdot$}}}
\pscircle[linewidth=1pt,fillstyle=solid,fillcolor=white](1.05,0){.15}
\rput(1.3,0){\textcolor{brown}{\hbox{\Large$\cdot$}}}
\pscircle[linewidth=1pt,fillstyle=solid,fillcolor=white](1.55,0){.15}
\rput(1.8,0){\textcolor{brown}{\hbox{\Large$\cdot$}}}
\pscircle[linewidth=1pt,fillstyle=solid,fillcolor=white](2.05,0){.15}
\rput(2.3,0){\textcolor{brown}{\hbox{\Large$\cdot$}}}
\pcline[linewidth=2pt,linecolor=brown]{->}(2.4,0)(2.8,0)
\pscircle[linewidth=1pt,fillstyle=solid,fillcolor=white](3,0){.2}
}
}
\rput(-5,0){\psarc[linecolor=blue, linewidth=2pt](0,-1){3.20}{22}{158}}
\rput(-3.5,0){\psarc[linecolor=blue, linewidth=2pt](0,0){1.5}{180}{360}}
\rput(-8,0){\PATTERN{1}}
\rput(-5,0){\PATTERN{1}}
\rput(-8,-.5){$p$}
\rput(-5.3,-.5){$r$}
\rput(-4.48,0){\psarc[linecolor=red, linestyle=dashed, linewidth=2pt]{<-}(0,-1.4){2.85}{42}{147}}
\rput(-4.23,0){\psarc[linecolor=red, linestyle=dashed, linewidth=2pt]{<-}(0,-2.23){3.15}{53}{132}}
\rput(-3.98,0){\psarc[linecolor=red, linestyle=dashed, linewidth=2pt]{<-}(0,-3.5){4}{65}{117}}
\rput(-3.02,0){\psarc[linecolor=red, linestyle=dashed, linewidth=2pt]{->}(0,.3){1.08}{205}{325}}
\rput(-2.8,0){\psarc[linecolor=red, linestyle=dashed, linewidth=2pt]{->}(0,.4){.86}{215}{313}}
\rput(-2.48,0){\psarc[linecolor=red, linestyle=dashed, linewidth=2pt]{->}(0,.48){.68}{232}{300}}
\rput(-1.7,0){$q$}
\pcline[linewidth=1pt]{>-<}(-4.5,0.3)(-5.15,1.6)
\rput(-5.5,1.7){$k$}
\pcline[linewidth=1pt]{>-<}(-2.5,-0.1)(-3.3,-.9)
\rput(-3.5,-1){$\rho$}
\rput(-1,0){\makebox(0,0)[lc]{$\displaystyle\sim\oint_{{\cal C}_{\cal D}^{(q)}}dE_{q,\bar q}(\bullet)
\frac{dq}{2\pi i\ y(q)}\frac{\d^\rho}{\d q^\rho}\Bigl(B(r,q)\frac{\d^k}{\d q^k}\bigl(B(p,q)\bigr)\Bigr),$}}
\rput(0,-1.5){\makebox(0,0)[lc]{$\displaystyle k\ge 0,\ \rho\ge 0
$}}
\end{pspicture}
}
\label{diagram*3*1}
\ee
Here the vertex $p$ must be outside all the starting points of dashed lines terminating at $q$ and can be
an external vertex. The vertex $r$ can be external if $k=0$.
\be
{\psset{unit=.8}
\begin{pspicture}(-8,-1)(10,2.5)
\newcommand{\PATTERN}[1]{%
{\psset{unit=#1}
\pscircle[linewidth=1pt,fillstyle=solid,fillcolor=white](0,0){.2}
\pcline[linewidth=2pt,linecolor=brown]{->}(0.2,0)(.7,0)
\rput(0.8,0){\textcolor{brown}{\hbox{\Large$\cdot$}}}
\pscircle[linewidth=1pt,fillstyle=solid,fillcolor=white](1.05,0){.15}
\rput(1.3,0){\textcolor{brown}{\hbox{\Large$\cdot$}}}
\pscircle[linewidth=1pt,fillstyle=solid,fillcolor=white](1.55,0){.15}
\rput(1.8,0){\textcolor{brown}{\hbox{\Large$\cdot$}}}
\pscircle[linewidth=1pt,fillstyle=solid,fillcolor=white](2.05,0){.15}
\rput(2.3,0){\textcolor{brown}{\hbox{\Large$\cdot$}}}
\pcline[linewidth=2pt,linecolor=brown]{->}(2.4,0)(3.4,0)
\rput(3.6,0){\textcolor{brown}{\hbox{$\bullet$}}}
\rput(4.1,0){\textcolor{brown}{\hbox{$\bullet$}}}
\rput(4.6,0){\textcolor{brown}{\hbox{$\bullet$}}}
\pcline[linewidth=2pt,linecolor=brown]{->}(5,0)(5.8,0)
\pscircle[linewidth=1pt,fillstyle=solid,fillcolor=white](6,0){.2}
}
}
\rput(-5,0){\psarc[linecolor=blue, linewidth=2pt](0,-1){3.20}{22}{158}}
\rput(-8,0){\PATTERN{1}}
\rput(-8,-.5){$p$}
\rput(-4.48,0){\psarc[linecolor=red, linestyle=dashed, linewidth=2pt]{<-}(0,-1.4){2.85}{42}{147}}
\rput(-4.23,0){\psarc[linecolor=red, linestyle=dashed, linewidth=2pt]{<-}(0,-2.23){3.15}{53}{132}}
\rput(-3.98,0){\psarc[linecolor=red, linestyle=dashed, linewidth=2pt]{<-}(0,-3.5){4}{65}{117}}
\rput(-2,-0.5){$q$}
\pcline[linewidth=1pt]{>-<}(-4.5,0.3)(-5.15,1.6)
\rput(-5.5,1.7){$k$}
\pcline[linewidth=2pt,linecolor=brown]{->}(-1.8,0)(-1,0)
\rput(-0.8,0){\textcolor{brown}{\hbox{\Large$\star$}}}
\rput(0,0){\makebox(0,0)[lc]{$\displaystyle\sim\oint_{{\cal C}_{\cal D}^{(q)}}dE_{q,\bar q}(\bullet)
\frac{dq}{2\pi i\ y(q)}\frac{\d^k}{\d q^k}\bigl(B(p,q)\bigr)dE_{\star,\bar{\star}}(q),\  k\ge 0.$}}
\end{pspicture}
}
\label{diagram*3*2}
\ee
Here the vertex $p$ must be outside all the starting points of dashed lines acting on the vertex $q$ and can be external.
\be
{\psset{unit=.8}
\begin{pspicture}(-5,-1)(5,1)
\rput(-4,0){
\rput(-.2,0){\textcolor{brown}{\hbox{$\bullet$}}}
\pcline[linewidth=2pt,linecolor=brown]{->}(0,0)(.8,0)
\psbezier[linewidth=2pt,linecolor=blue](1.1,0.1)(2.4,1.5)(2.4,-1.5)(1.1,-0.1)
\pscircle[linewidth=1pt,fillstyle=solid,fillcolor=white](1,0){.2}
\rput(.9,-.5){$q$}
}
\rput(-0.5,0){\makebox(0,0)[lc]{$\displaystyle\sim\oint_{{\cal C}_{\cal D}^{(q)}}dE_{q,\bar q}(\bullet)
\frac{dq}{2\pi i\ y(q)}B(q,\bar q)$}}
\end{pspicture}
}
\label{diagram*3*11}
\ee
\be
{\psset{unit=.8}
\begin{pspicture}(-5,-1)(5,1)
\rput(-4,0){
\rput(-.2,0){\textcolor{brown}{\hbox{$\bullet$}}}
\pcline[linewidth=2pt,linecolor=brown]{->}(0,0)(.8,0)
\pcline[linewidth=2pt,linecolor=brown]{->}(1.2,0)(1.8,0)
\rput(2,0){\textcolor{brown}{\hbox{\Large$\star$}}}
\rput(2.3,0){\textcolor{brown}{\hbox{$\bullet$}}}
\pcline[linewidth=2pt,linecolor=brown]{->}(2.5,0)(2.85,0)
\pcline[linewidth=2pt,linecolor=brown]{->}(3.15,0)(3.7,0)
\psbezier[linewidth=2pt,linecolor=blue](1.1,0.1)(1.8,1)(2.2,1)(2.9,0.1)
\pscircle[linewidth=1pt,fillstyle=solid,fillcolor=white](1,0){.2}
\pscircle[linewidth=1pt,fillstyle=solid,fillcolor=black](3,0.1){.15}
\pscircle[linewidth=1pt,fillstyle=solid,fillcolor=white](3,-0.1){.15}
\rput(1,-.5){$q$}
\rput(3,.5){\textcolor{brown}{\hbox{\Large$\ast$}}}
}
\rput(0.5,0){\makebox(0,0)[lc]{$\displaystyle\sim\oint_{{\cal C}_{\cal D}^{(q)}}dE_{q,\bar q}(\bullet)
\frac{dq}{2\pi i\ y(q)}B(q,\ast)dE_{\star,\bar{\star}}(q)$}}
\end{pspicture}
}
\label{diagram*3*3}
\ee
\be
{\psset{unit=.8}
\begin{pspicture}(-5,-1)(5,1)
\rput(-4,0){
\rput(-.2,0){\textcolor{brown}{\hbox{$\bullet$}}}
\pcline[linewidth=2pt,linecolor=brown]{->}(0,0)(.8,0)
\pcline[linewidth=2pt,linecolor=brown]{->}(1.1,0.1)(1.7,0.7)
\pcline[linewidth=2pt,linecolor=brown]{->}(1.1,-0.1)(1.7,-0.7)
\rput(1.9,0.9){\textcolor{brown}{\hbox{\Large$\star$}}}
\rput(1.9,-0.9){\textcolor{brown}{\hbox{\Large$\ast$}}}
\pscircle[linewidth=1pt,fillstyle=solid,fillcolor=white](1,0){.2}
\rput(1,-.5){$q$}
}
\rput(-1,0){\makebox(0,0)[lc]{$\displaystyle\sim\oint_{{\cal C}_{\cal D}^{(q)}}dE_{q,\bar q}(\bullet)
\frac{dq}{2\pi i\ y(q)}dE_{\ast,\bar{\ast}}(q)dE_{\star,\bar{\star}}(q)$}}
\end{pspicture}
}
\label{diagram*3*4}
\ee

{\em The vertices with two adjacent solid lines}
\be
{\psset{unit=.8}
\begin{pspicture}(-8,-1.5)(8,1.5)
\newcommand{\PATTERN}[1]{%
{\psset{unit=#1}
\pcline[linewidth=2pt,linecolor=brown]{->}(0.2,0)(.7,0)
\rput(0.8,0){\textcolor{brown}{\hbox{\Large$\cdot$}}}
\pscircle[linewidth=1pt,fillstyle=solid,fillcolor=white](1.05,0){.15}
\rput(1.3,0){\textcolor{brown}{\hbox{\Large$\cdot$}}}
\pscircle[linewidth=1pt,fillstyle=solid,fillcolor=white](1.55,0){.15}
\rput(1.8,0){\textcolor{brown}{\hbox{\Large$\cdot$}}}
\pscircle[linewidth=1pt,fillstyle=solid,fillcolor=white](2.05,0){.15}
\rput(2.3,0){\textcolor{brown}{\hbox{\Large$\cdot$}}}
\pcline[linewidth=2pt,linecolor=brown]{->}(2.4,0)(2.8,0)
}
}
\rput(-3.5,0){\psarc[linecolor=blue, linewidth=2pt](0,0){1.5}{180}{360}}
\rput(-8,0){\PATTERN{1}}
\rput(-5,0){\PATTERN{1}}
\rput(-5.3,-.5){$r$}
\pscircle[linewidth=1pt,fillstyle=solid,fillcolor=white](-5,0){.2}
\pscircle[linewidth=1pt,fillstyle=solid,fillcolor=black](-2,0){.2}
\rput(-4.48,0){\psarc[linecolor=red, linestyle=dashed, linewidth=2pt]{<-}(0,-1.4){2.85}{42}{147}}
\rput(-4.23,0){\psarc[linecolor=red, linestyle=dashed, linewidth=2pt]{<-}(0,-2.23){3.15}{53}{132}}
\rput(-3.98,0){\psarc[linecolor=red, linestyle=dashed, linewidth=2pt]{<-}(0,-3.5){4}{65}{117}}
\rput(-3.02,0){\psarc[linecolor=red, linestyle=dashed, linewidth=2pt]{->}(0,.3){1.08}{205}{325}}
\rput(-2.8,0){\psarc[linecolor=red, linestyle=dashed, linewidth=2pt]{->}(0,.4){.86}{215}{313}}
\rput(-2.48,0){\psarc[linecolor=red, linestyle=dashed, linewidth=2pt]{->}(0,.48){.68}{232}{300}}
\rput(-1.5,0){$q$}
\pcline[linewidth=1pt]{>-<}(-4.5,0.3)(-5.15,1.6)
\rput(-5.5,1.7){$k$}
\pcline[linewidth=1pt]{>-<}(-2.5,-0.1)(-3.3,-.9)
\rput(-3.5,-1){$\rho$}
\rput(-1,0){\makebox(0,0)[lc]{$\displaystyle\sim\oint_{{\cal C}_{\cal D}^{(q)}}dE_{q,\bar q}(\bullet)
\frac{dq}{2\pi i\ y(q)}\frac{\d^\rho}{\d q^\rho}\Bigl(B(r,q)y^{(k)}(q)\Bigr),$}}
\rput(0,-1.5){\makebox(0,0)[lc]{$\displaystyle k> 0,\ \rho\ge 0.
$}}
\end{pspicture}
}
\label{diagram*2*1}
\ee
Here, note that $k$ {\em must be greater than zero}, so the point $r$ cannot be external,
\be
{\psset{unit=.8}
\begin{pspicture}(-8,-1.5)(8,1.5)
\rput(-2,0){
\rput(-3.3,0){\textcolor{brown}{\hbox{$\bullet$}}}
\pcline[linewidth=2pt,linecolor=brown]{->}(-3,0)(-2.2,0)
\pcline[linewidth=2pt,linecolor=brown]{->}(-1.8,0)(-1,0)
\pscircle[linewidth=1pt,fillstyle=solid,fillcolor=black](-2,0){.2}
\rput(-.7,0){\textcolor{brown}{\hbox{\Large$\star$}}}
\rput(-4.48,0){\psarc[linecolor=red, linestyle=dashed, linewidth=2pt]{<-}(0,-1.4){2.85}{42}{70}}
\rput(-4.23,0){\psarc[linecolor=red, linestyle=dashed, linewidth=2pt]{<-}(0,-2.23){3.15}{53}{75}}
\rput(-3.98,0){\psarc[linecolor=red, linestyle=dashed, linewidth=2pt]{<-}(0,-3.5){4}{65}{85}}
\rput(-2,-0.5){$q$}
\pcline[linewidth=1pt]{>-<}(-3.5,0.3)(-3.15,1.35)
\rput(-2.9,1.4){$k$}
}
\rput(-1,0){\makebox(0,0)[lc]{$\displaystyle\sim\oint_{{\cal C}_{\cal D}^{(q)}}dE_{q,\bar q}(\bullet)
\frac{dq}{2\pi i\ y(q)}y^{(k)}(q)dE_{\star,\bar{\star}}(q),\ k> 0.$}}
\end{pspicture}
}
\label{diagram*2*2}
\ee
\be
{\psset{unit=.8}
\begin{pspicture}(-8,-1)(10,2.5)
\newcommand{\PATTERN}[1]{%
{\psset{unit=#1}
\pscircle[linewidth=1pt,fillstyle=solid,fillcolor=white](0,0){.2}
\pcline[linewidth=2pt,linecolor=brown]{->}(0.2,0)(.7,0)
\rput(0.8,0){\textcolor{brown}{\hbox{\Large$\cdot$}}}
\pscircle[linewidth=1pt,fillstyle=solid,fillcolor=white](1.05,0){.15}
\rput(1.3,0){\textcolor{brown}{\hbox{\Large$\cdot$}}}
\pscircle[linewidth=1pt,fillstyle=solid,fillcolor=white](1.55,0){.15}
\rput(1.8,0){\textcolor{brown}{\hbox{\Large$\cdot$}}}
\pscircle[linewidth=1pt,fillstyle=solid,fillcolor=white](2.05,0){.15}
\rput(2.3,0){\textcolor{brown}{\hbox{\Large$\cdot$}}}
\pcline[linewidth=2pt,linecolor=brown]{->}(2.4,0)(3.4,0)
\rput(3.6,0){\textcolor{brown}{\hbox{$\bullet$}}}
\rput(4.1,0){\textcolor{brown}{\hbox{$\bullet$}}}
\rput(4.6,0){\textcolor{brown}{\hbox{$\bullet$}}}
\pcline[linewidth=2pt,linecolor=brown]{->}(5,0)(5.8,0)
\pscircle[linewidth=1pt,fillstyle=solid,fillcolor=white](6,0){.2}
}
}
\rput(-5,0){\psarc[linecolor=blue, linewidth=2pt](0,-1){3.20}{22}{158}}
\rput(-8,0){\PATTERN{1}}
\rput(-8,-.5){$p$}
\rput(-4.48,0){\psarc[linecolor=red, linestyle=dashed, linewidth=2pt]{<-}(0,-1.4){2.85}{42}{147}}
\rput(-4.23,0){\psarc[linecolor=red, linestyle=dashed, linewidth=2pt]{<-}(0,-2.23){3.15}{53}{132}}
\rput(-3.98,0){\psarc[linecolor=red, linestyle=dashed, linewidth=2pt]{<-}(0,-3.5){4}{65}{117}}
\rput(-2,-0.5){$q$}
\pcline[linewidth=1pt]{>-<}(-4.5,0.3)(-5.15,1.6)
\rput(-5.5,1.7){$k$}
\psbezier[linewidth=2pt,linecolor=red,linestyle=dashed]{->}(-1.8,0.2)(-.5,1.5)(-.5,-1.5)(-1.8,-0.2)
\rput(0,0){\makebox(0,0)[lc]{$\displaystyle\sim\oint_{{\cal C}_{\cal D}^{(q)}}dE_{q,\bar q}(\bullet)
\frac{dq}{2\pi i\ y(q)}\frac{\d^{k+1}}{\d q^{k+1}}B(p,q),\  k\ge 0.$}}
\end{pspicture}
}
\label{diagram*2*3}
\ee
Here the vertex $p$ must be outside all the starting points of dashed lines acting on the vertex $q$ and can be external.
\be
{\psset{unit=.8}
\begin{pspicture}(-5,-1)(5,1)
\rput(-4,0){
\rput(-.2,0){\textcolor{brown}{\hbox{$\bullet$}}}
\pcline[linewidth=2pt,linecolor=brown]{->}(0,0)(.8,0)
\pcline[linewidth=2pt,linecolor=brown]{->}(1.2,0)(1.8,0)
\rput(2,0){\textcolor{brown}{\hbox{\Large$\star$}}}
\rput(2.3,0){\textcolor{brown}{\hbox{$\bullet$}}}
\pcline[linewidth=2pt,linecolor=brown]{->}(2.5,0)(2.85,0)
\pcline[linewidth=2pt,linecolor=brown]{->}(3.15,0)(3.7,0)
\psbezier[linewidth=2pt,linecolor=red,linestyle=dashed]{->}(1.1,0.1)(1.8,1)(2.2,1)(2.85,0.1)
\pscircle[linewidth=1pt,fillstyle=solid,fillcolor=white](1,0){.2}
\pscircle[linewidth=1pt,fillstyle=solid,fillcolor=black](3,0.1){.15}
\pscircle[linewidth=1pt,fillstyle=solid,fillcolor=white](3,-0.1){.15}
\rput(1,-.5){$q$}
}
\rput(0.5,0){\makebox(0,0)[lc]{$\displaystyle\sim\oint_{{\cal C}_{\cal D}^{(q)}}dE_{q,\bar q}(\bullet)
\frac{dq}{2\pi i\ y(q)}dE_{\star,\bar{\star}}(q)$}}
\end{pspicture}
}
\label{diagram*2*4}
\ee

{\em The vertex with one adjacent solid line}
\be
{\psset{unit=.8}
\begin{pspicture}(-8,-1)(8,1.5)
\rput(-2,0){
\rput(-3.3,0){\textcolor{brown}{\hbox{$\bullet$}}}
\pcline[linewidth=2pt,linecolor=brown]{->}(-3,0)(-2.2,0)
\pscircle[linewidth=1pt,fillstyle=solid,fillcolor=black](-2,0){.2}
\rput(-4.48,0){\psarc[linecolor=red, linestyle=dashed, linewidth=2pt]{<-}(0,-1.4){2.85}{42}{70}}
\rput(-4.23,0){\psarc[linecolor=red, linestyle=dashed, linewidth=2pt]{<-}(0,-2.23){3.15}{53}{75}}
\rput(-3.98,0){\psarc[linecolor=red, linestyle=dashed, linewidth=2pt]{<-}(0,-3.5){4}{65}{85}}
\rput(-2,-0.5){$q$}
\pcline[linewidth=1pt]{>-<}(-3.5,0.3)(-3.15,1.35)
\rput(-2.9,1.4){$k$}
\psbezier[linewidth=2pt,linecolor=red,linestyle=dashed]{->}(-1.8,0.2)(-.5,1.5)(-.5,-1.5)(-1.8,-0.2)
}
\rput(-1,0){\makebox(0,0)[lc]{$\displaystyle\sim\oint_{{\cal C}_{\cal D}^{(q)}}dE_{q,\bar q}(\bullet)
\frac{dq}{2\pi i\ y(q)}y^{(k+1)}(q),\ k\ge 0.$}}
\end{pspicture}
}
\label{diagram*1*1}
\ee

When calculating $W_{k,l}(p_1,\dots,p_s)$ the order of integration contours (previously the order of
taking residues at the branching points in the case of the Hermitian one- and
two-matrix models) is prescribed by the ordering of vertices in the tree subgraph: the
closer is the vertex to the root, the more outer is the integration
contour. In contrast to the Hermitian matrix model case, the integration cannot be
reduced to taking residues at the branch points only; all
internal integrations can be nevertheless reduced to sums of
residues, but these sums may now include residues at zeros of the
additional polynomial $M(p)$ on the nonphysical sheet and, possibly,
at the point $\infty_-$.

As in the Hermitian matrix-model case,
we use the $H$-operator constructed below
to invert the action of the loop insertion operator and
obtain the expression for the free energy itself.

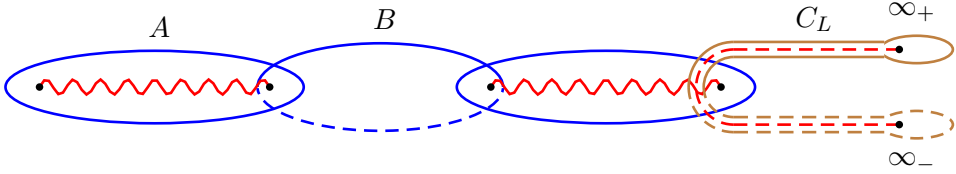
\begin{figure}[tb]
{\psset{unit=1}
\begin{pspicture}(-8,-1.3)(8,1.3)
\newcommand{\OVAL}{%
\psellipse[linecolor=blue, linewidth=1pt](0,0)(2,0.5)
\parametricplot[linecolor=red,linewidth=1pt]{-1530}{1530}{0.001 t mul 0.1 t cos mul}
\pscircle*(-1.53,0){0.05}
\pscircle*(1.53,0){0.05}
}
\newcommand{\CL}{%
\psellipse[linecolor=brown, linewidth=1pt](2.43,0.5)(.5,0.2)
\psellipse[linecolor=brown, linestyle=dashed, linewidth=1pt](2.43,-0.5)(.5,0.2)
\psframe[linecolor=white, fillstyle=solid, fillcolor=white](1.5,-1)(2,1)
\psarc[linecolor=brown, linewidth=1pt](0,0){0.4}{90}{180}
\psarc[linecolor=red, linestyle=dashed, linewidth=1pt](0,0){0.5}{90}{270}
\psarc[linecolor=brown, linewidth=1pt](0,0){0.6}{90}{180}
\psarc[linecolor=brown, linestyle=dashed, linewidth=1pt](0,0){0.4}{180}{270}
\psarc[linecolor=brown, linestyle=dashed, linewidth=1pt](0,0){0.6}{180}{270}
\pcline[linecolor=brown, linewidth=1pt](0,0.4)(2,0.4)
\pcline[linecolor=brown, linewidth=1pt](0,0.6)(2,0.6)
\pcline[linecolor=red, linestyle=dashed, linewidth=1pt](0,0.5)(2.2,0.5)
\pcline[linecolor=brown, linestyle=dashed, linewidth=1pt](0,-0.4)(2,-0.4)
\pcline[linecolor=brown, linestyle=dashed, linewidth=1pt](0,-0.6)(2,-0.6)
\pcline[linecolor=red, linestyle=dashed, linewidth=1pt](0,-0.5)(2.2,-0.5)
\pscircle*(2.2,0.5){0.05}
\pscircle*(2.2,-0.5){0.05}
}
\newcommand{\BCYCLE}{%
\psellipse[linecolor=blue, linewidth=1pt](0,0)(1.65,0.6)
\psframe[linecolor=white, fillstyle=solid, fillcolor=white](-1.7,-1)(1.7,0)
\psclip{\psframe[linecolor=white, fillstyle=solid, fillcolor=white](-1.7,-1)(1.7,0)}
\psellipse[linecolor=blue, linestyle=dashed, linewidth=1pt](0,0)(1.65,0.6)
\endpsclip
}
\rput(-1,0){\BCYCLE}
\rput(-4,0){\OVAL}
\rput(2,0){\OVAL}
\rput(3.7,0){\CL}
\rput(-4,0.8){\makebox(0,0){ $A$}}
\rput(-1,0.9){\makebox(0,0){ $B$}}
\rput(4.7,0.9){\makebox(0,0){ $C_L$}}
\rput(6,1){\makebox(0,0){ $\infty_+$}}
\rput(6,-1){\makebox(0,0){ $\infty_-$}}
\end{pspicture}
}
\caption{Cuts in the $\lambda$-, or ``eigenvalue," plane for the
planar limit of the $\beta$-model (coinciding with the
planar limit of the Hermitian one-matrix model). The eigenvalues are supposed to be
located ``on" the cuts (curly lines). We add the logarithmic
cut between two copies of the infinity on two sheets of the
hyperelliptic Riemann surface in order to calculate the derivative
w.r.t. the variable $t_0$; $C_L$ is the corresponding regularized integration contour.}
\label{fi:cuts}
\end{figure}

\section{Inverting the loop insertion operator. Free energy  \label{ss:freeBeta}}

\paragraph{The $H$-operator. \label{ss:H}}

We now introduce the operator (the functional on 1-forms)
that is inverse to loop insertion operator (\ref{X.6}). For a form $df(x)$,
let\footnote{This definition works well when acting on 1-forms regular at infinities. Otherwise (say, in the case of
$W_0(p)$), the integral in the third term must be regularized by replacing it by the integral along the contour
$C_L$ depicted in Fig.~\ref{fi:cuts}.}
\be
H_x df(x)=\frac12\res_{\infty_+}V(x)df(x) -\frac12\res_{\infty_-}V(x)df(x)
-t_0\int_{\infty_-}^{\infty_+}df(x) -\sum_{i=1}^{n-1}S_i\oint_{B_i}df(x) .
\label{H}
\ee
The arrangement of the integration contours is as in Fig.~\ref{fi:cuts}.

The $H$-operator, in contrast to the loop insertion operator, therefore
reduce by one the degree of the (symmetrized) form associated with an $s$-point correlation function
\be
\label{H-Omega}
H:\, \Omega^s_{1,0}\mapsto \Omega^{s-1}_{1,0}.
\ee
We calculate the action of $H$ on the Bergmann
bidifferential $B(x,q)$ using again the Riemann bilinear identities.
We first note that as $B(x,q)=\d_x dE_{x,q_0}(q)$, we can evaluate residues at infinities
by parts. Then, since $dE_{x,q_0}(q)$ is regular at infinities, we substitute $2y(x)+2t_0/x$ for $V'(x)$ as $x\to\infty_+$
and $-2y(x)+2t_0/x$ for $V'(x)$ as $x\to\infty_-$ thus obtaining
\bea
&&-\res_{\infty_+}\left(y(x)+\frac{t_0}{x}\right)dE_{x,q_0}(q)dx+\res_{\infty_-}\left(-y(x)+\frac{t_0}{x}\right)dE_{x,q_0}(q)dx
\nonumber
\\
&&\qquad \Bigl.-t_0dE_{x,q_0}(q)\Bigr|_{x=\infty_-}^{x=\infty_+}-\sum_{i=1}^{n-1}S_i\oint_{B_i}B(q,x).
\label{Z.1}
\eea
Whereas the cancelation of terms containing $t_0$ is obvious, it remains only to take the combination of residues
at infinities involving $y(x)$. For this, we cut the surface along $A$- and $B$-cycles taking into account the residue at
$x=q$. The boundary integrals on two sides of the cut at $B_i$ then differ by $dE_{x,q_0}(q)-dE_{x+\oint_{A_i},q_0}(q)=0$,
while the integrals on the two sides of the cut at $A_i$ differ by  $dE_{x,q_0}(q)-dE_{x+\oint_{B_i},q_0}(q)=\oint_{B_i}B(q,x)$,
and the boundary term therefore becomes
$$
\sum_{i=1}^{n-1}\oint_{A_i}y(x)dx\oint_{B_i}B(q,\xi),
$$
which exactly cancels the last term in (\ref{Z.1}). Only the contribution from the pole at $x=q$ then survives, and this
contribution is just $-y(q)$. We have therefore proved that
\be
H_x\cdot B(x,q)=-y(q)dq.
\label{HB}
\ee

Let us now consider the action of the operator
$H_x$ given by (\ref{H}) on $W_{k,l}(x)$ subsequently evaluating the action of loop insertion operator on the
result. Due to the commutation relations between $H_x$ and $\partial/\partial V(x)$,
$$
H \circ \partial/\partial V-\partial/\partial V\circ H=\hbox{Id}\quad \hbox{modulo the kernels},
$$
we have,
\be
\frac{\d}{\d V(p)}\left(H_x\cdot W_{k,l}(x)\right)=W_{k,l}(p)+H_x\cdot W_{k,l}(x,p).
\label{Z.2-1}
\ee
For the second term, due to the symmetry of $W_{k,l}(p,q)$, we may choose the point $p$ to be the root of the tree subgraphs. Then,
the operator $H_x$ always acts on $B(x,\xi)$ (or, possibly, on its derivatives w.r.t. $\xi$)
where $\xi$ are some integration variables of internal vertices.

Let us consider the action of $\d/\d V(p)$ on the elements of the Feynman diagram technique in Sec.~\ref{ss:rules}.
Here we have three different cases; in all of them the variable $p$ is external.
\begin{itemize}
    \item When acting on the arrowed propagator followed by a (white or black) vertex, we use the standard
variational relation  in (\ref{variation}) below, where, again, $F(\eta)$ is any combination of elements of the diagrammatic technique,
\be
{\psset{unit=.8}
\begin{pspicture}(-4,-1)(4,1)
\rput(-4,0){
\rput(-1,0){\makebox(0,0)[cc]{$\displaystyle \frac{\partial}{\partial V(p)}$}}
\pcline[linewidth=2pt,linecolor=brown]{->}(0,0)(.8,0)
\pscircle[linewidth=1pt,fillstyle=solid,fillcolor=white](1,0){.2}
\rput(0,-.4){\makebox(0,0)[cc]{$q$}}
\rput(1,-.5){\makebox(0,0)[cc]{$\eta$}}
\rput(1.2,0){\makebox(0,0)[lc]{$\Bigl\{F(\eta)\Bigr\}$}}
}
\rput(-.5,0){\makebox(0,0)[cc]{$=$}}
\rput(0,0){
\rput(-0.2,0.8){\makebox(0,0)[cc]{$p$}}
\pcline[linewidth=2pt,linecolor=brown]{->}(0,0)(.8,0)
\psbezier[linewidth=2pt,linecolor=blue](0,0.8)(0.8,.7)(.8,.7)(.95,0.1)
\pcline[linewidth=2pt,linecolor=brown]{->}(1.1,0)(1.8,0)
\pscircle[linewidth=1pt,fillstyle=solid,fillcolor=white](1,0){.2}
\pscircle[linewidth=1pt,fillstyle=solid,fillcolor=white](2,0){.2}
\rput(0,-.4){\makebox(0,0)[cc]{$q$}}
\rput(1,-.5){\makebox(0,0)[cc]{$\xi$}}
\rput(2,-.5){\makebox(0,0)[cc]{$\eta$}}
\rput(2.2,0){\makebox(0,0)[lc]{$\Bigl\{\displaystyle F(\eta)\Bigr\}$}}
}
\end{pspicture}
}
\label{variation}
\ee
\item When acting on nonarrowed internal propagator $\frac{\d^k}{\d q^k}B(r,q)$, $r> q$,
$k\ge 0$, we apply the relation
\be
{\psset{unit=.8}
\begin{pspicture}(-4,-1)(4,1)
\rput(-4,0){
\rput(-2,0){\makebox(0,0)[cc]{$\displaystyle \frac{\partial}{\partial V(p)}\frac{\d^{k}}{\d q^{k}}$}}
\psbezier[linewidth=2pt,linecolor=blue](0,0)(0.7,-.7)(.8,-.7)(1.5,0)
\pcline[linewidth=2pt,linecolor=brown]{->}(0,0)(.5,0)
\rput(.75,0){\textcolor{brown}{\hbox{$\bullet$}}}
\pcline[linewidth=2pt,linecolor=brown]{->}(1,0)(1.5,0)
\rput(0,-.4){\makebox(0,0)[cc]{$r$}}
\rput(1.5,-.4){\makebox(0,0)[cc]{$q$}}
}
\rput(-1,0){\makebox(0,0)[cc]{$=$}}
\rput(-0.5,0){
\psbezier[linewidth=2pt,linecolor=blue](0,0)(1,-1)(1.5,-1)(2.5,0)
\pcline[linewidth=2pt,linecolor=brown]{->}(0,0)(.7,0)
\rput(1.4,0){\makebox(0,0)[rc]{$\frac{\d^{k}}{\d q^{k}}$}}
\pcline[linewidth=2pt,linecolor=brown]{->}(1.5,0)(2.3,0)
\psbezier[linewidth=2pt,linecolor=blue](0,0.8)(1.5,0.8)(1.5,0.8)(2.5,0)
\pscircle[linewidth=1pt,fillstyle=solid,fillcolor=white](2.5,0){.2}
\rput(0,-.4){\makebox(0,0)[cc]{$r$}}
\rput(1.5,-.4){\makebox(0,0)[cc]{$q$}}
\rput(-0.2,0.8){\makebox(0,0)[cc]{$p$}}
}
\end{pspicture}
}
\label{Y.11}
\ee
{\em without} subsequent expanding the action of the derivative $\frac{\d^k}{\d q^k}$
into a sum of diagrams.
\item Eventually, when acting on the factors $y^{(k)}(q)$ standing at the black vertices, using relation (\ref{dydV}) we obtain
from the diagrams (\ref{diagram*2*1}), (\ref{diagram*2*2}), and (\ref{diagram*1*1}) the respective
diagrams (\ref{diagram*3*1}), (\ref{diagram*3*2}), and (\ref{diagram*2*3}).
\end{itemize}

We now consider the {\em inverse} action of the $H$-operator in all three cases.

In the first case where it exists an outgoing arrowed propagator $dE_{\eta,\bar \eta}(\xi)$
we can push the integration contour for $\xi$ through the one for $p$;
the only contribution comes from the pole at $\xi=p$ (with the {\em opposite} sign due to the choice of contour directions in
Fig.~\ref{fi:cuts}). Graphically, we have
\be
{\psset{unit=.8}
\begin{pspicture}(-4,-1)(4,1)
\rput(1.5,0){
\pcline[linewidth=2pt,linecolor=brown]{->}(0,0)(.8,0)
\pscircle[linewidth=1pt,fillstyle=solid,fillcolor=white](1,0){.2}
\rput(0,-.4){\makebox(0,0)[cc]{$q$}}
\rput(1,-.5){\makebox(0,0)[cc]{$\eta$}}
\rput(1.2,0){\makebox(0,0)[lc]{$\Bigl\{F(\eta)\Bigr\}$}}
}
\rput(.5,0){\makebox(0,0)[cc]{$=$}}
\rput(1,0){\makebox(0,0)[cc]{$-$}}
\rput(-4,0){
\rput(-0.3,0.8){\makebox(0,0)[cc]{$H_\cdot$}}
\pcline[linewidth=2pt,linecolor=brown]{->}(0,0)(.8,0)
\psbezier[linewidth=2pt,linecolor=blue](0,0.8)(0.8,.7)(.8,.7)(.95,0.1)
\pcline[linewidth=2pt,linecolor=brown]{->}(1.1,0)(1.8,0)
\pscircle[linewidth=1pt,fillstyle=solid,fillcolor=white](1,0){.2}
\pscircle[linewidth=1pt,fillstyle=solid,fillcolor=white](2,0){.2}
\rput(0,-.4){\makebox(0,0)[cc]{$q$}}
\rput(1,-.5){\makebox(0,0)[cc]{$\xi$}}
\rput(2,-.5){\makebox(0,0)[cc]{$\eta$}}
\rput(2.2,0){\makebox(0,0)[lc]{$\Bigl\{\displaystyle F(\eta)\Bigr\}$}}
}
\end{pspicture}
}
\label{chopping1}
\ee

In the second case, the vertex $\xi$ in (\ref{Y.11})
is an innermost vertex (i.e., there is no arrowed edges coming out of it). The 1-form $y(\xi)d\xi$ arising under the action of $H$
(\ref{HB}) cancels the corresponding form in the integration expression,
the expression becomes regular at the branching point and the residue vanishes.
Graphically, we have for $k\ge 0$,
\be
{\psset{unit=.8}
\begin{pspicture}(-4,-1)(4,1)
\rput(1,0){\makebox(0,0)[cc]{$=0.$}}
\rput(-2.5,0){
\psbezier[linewidth=2pt,linecolor=blue](0,0)(1,-1)(1.5,-1)(2.5,0)
\pcline[linewidth=2pt,linecolor=brown]{->}(0,0)(.7,0)
\rput(1.4,0){\makebox(0,0)[rc]{$\frac{\d^{k}}{\d q^{k}}$}}
\pcline[linewidth=2pt,linecolor=brown]{->}(1.5,0)(2.3,0)
\psbezier[linewidth=2pt,linecolor=blue](0,0.8)(1.5,0.8)(1.5,0.8)(2.5,0)
\pscircle[linewidth=1pt,fillstyle=solid,fillcolor=white](2.5,0){.2}
\rput(0,-.4){\makebox(0,0)[cc]{$r$}}
\rput(1.5,-.4){\makebox(0,0)[cc]{$q$}}
\rput(-0.3,0.8){\makebox(0,0)[cc]{$H_\cdot$}}
}
\end{pspicture}
}
\label{chopping2}
\ee

Eventually, in the third case,
the action of the $H$-operator just erases the new $B$-propagator with the
external variable $p$ and restore the term $y^{(k)}(q)$, and from the diagrams
(\ref{diagram*3*1}), (\ref{diagram*3*2}), and (\ref{diagram*2*3}) we return to the
respective initial diagrams (\ref{diagram*2*1}), (\ref{diagram*2*2}), and (\ref{diagram*1*1})
changing back the color of vertices from white to black. Note that in this case we have the
plus sign upon the action of the operator $H$.

For $H_x\cdot W_{k,l}(x,p)=H_x\cdot \frac{\d}{\d V(x)}W_{k,l}(p)$, we obtain that for each
solid arrowed edge, on which the action of
${\d}/{\d V(x)}$ produces the new (white) vertex, the inverse action of $H_x$ gives the factor $-1$,
on each nonarrowed edge,
on which the action of ${\d}/{\d V(x)}$ produces the new vertex accordingly to (\ref{Y.11}),
the inverse action of $H_x$ just gives zero, and at each black vertex, at which the action of the loop
insertion operator
changes the color to white and adds a new $B$-propagator, the inverse action of $H_x$ gives the factor $+1$.

As the total number
of arrowed edges coincides with the total number of vertices and the contributions of black vertices are opposite to the
contributions of arrowed edges, the total factor on which the diagram is multiplied is minus the number of white
vertices, which is exactly $2k+l-1$ for {\em any} graph contributing to $W_{k,l}(p)$. (For the $s$-point correlation function
$W_{k,l}(p_1,\dots,p_s)$ this number is $2k+l+s-2$.) We then have
$$
H_x\cdot W_{k,l}(x,p)=-(2k+l-1)W_k(p)
$$
and, combining with (\ref{Z.2-1}), we obtain
\be
\frac{\d}{\d V(p)}\left(H_x\cdot W_{k,l}(x)\right)=-(2-2k-l)\frac{\d}{\d V(p)}{\cal F}_{k,l}.
\label{Z.3-1}
\ee
We therefore obtain the final answer for the  {\sl free energy}:
\be
{\cal F}_{k,l}=\frac{1}{2k+l-2}H_x\cdot W_{k,l}(x),
\label{fin1}
\ee
which enables one to calculate {\em all} the terms
${\cal F}_{k,l}$ except the contribution at $k=1,l=0$ (which is the torus approximation
in the Hermitian one-matrix model calculated in~\cite{Ch1}) and the second-order correction in $\zeta$ (the term ${\cal F}_{0,2}$).
The term ${\cal F}_{0,2}$ was calculated in~\cite{ChEy-b}, in the next subsection we present the answer.

To calculate {\em all other} terms of the free energy expansion
we only need to introduce the {\bf new vertex} \hbox{$\circ$\hskip-0.85ex$\cdot$}
at which we place the (nonlocal) integral term
$\oint_{{\cal C}^{(\xi)}_{\cal D}}\frac{d\xi}{2\pi i}\frac{\int_{\bar \xi}^\xi y(s)ds}{2y(\xi)}$.
{\sl To obtain ${\cal F}_{g,l}$ in the above diagrammatic technique we merely remove the
root of the tree and replace the first internal vertex (which is always white) by this new (two-valent) vertex, as shown in examples below.}

In the Hermitian matrix model case, it was possible to perform the integration for
$\int_{\bar\xi}^\xi 2y(s)ds$ through the branch
point $\mu_\alpha$ in the vicinity of the $\alpha$th branch point; here it is no more the case and we must consider global
integrations. Note, however, that we must introduce nonlocal terms {\em only} for the very last, outermost, integration;
all internal integrations can be performed by taking residues at branch points and at the zeros of the polynomial $M(p)$.

For example, the first correction to the free energy is
\be
\hbox{
{\psset{unit=.8}
\begin{pspicture}(-3,-1)(3,1)
\rput(0.5,0){\makebox(0,0)[rc]{$-1\cdot {\cal F}_{0,1}=$}}
\pscircle[linewidth=1pt, fillstyle=solid, fillcolor=white](1,0){0.2}
\rput(1,0){\makebox(0,0){\tcg{$\bullet$}}}
\psbezier[linewidth=2pt,linecolor=red,linestyle=dashed]{->}(1.2,0.2)(2.5,1.5)(2.5,-1.5)(1.2,-0.2)
\end{pspicture}
}
}
\label{F01}
\ee
By integration by parts (in which we make substitution at $q$ and $\bar q$, which cancels the factor $2$ in the
denominator), we obtain
$$
{\cal F}_{0,1}=-\oint_{{\cal C}_{\cal D}^{(q)}}\frac{\int^q_{\bar q} y}{2y(q)}y'(q)\frac{dq}{2\pi i}=
\oint_{{\cal C}_{\cal D}^{(q)}}y(q)\log y(q)\frac{dq}{2\pi i},
$$
i.e., we come to the semiclassical Dyson term (the entropy). Other lower-order terms of the free-energy
expansion may also have a physical interpretation, as we demonstrate on the example of the term ${\mathcal F}_{0,2}$.

For the term ${\mathcal F}_{1,1}$ we have
$$
{\psset{unit=.8}
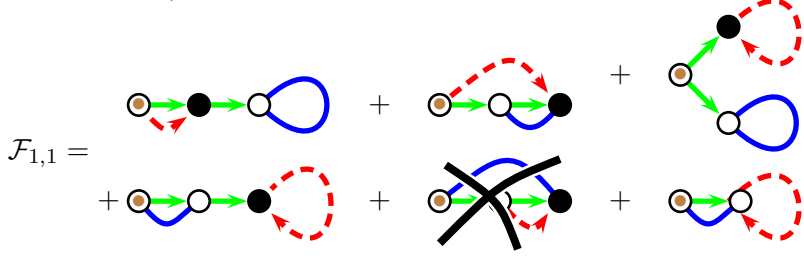
\begin{pspicture}(8,-2)(-4,2)
\rput(-6.5,0){\textcolor{black}{${\mathcal F}_{1,1}=$}}
\rput(0,0.8){
\psbezier[linewidth=2pt,linecolor=red,linestyle=dashed]{->}(-5,0)(-4.5,-.6)(-4.5,-.4)(-4.2,-0.2)
\pcline[linewidth=2pt,linecolor=green]{->}(-5,0)(-4.2,0)
\pscircle[linewidth=1pt, fillstyle=solid, fillcolor=white](-5,0){0.2}
\rput(-5,0){\makebox(0,0){\tcg{$\bullet$}}}
\pcline[linewidth=2pt,linecolor=green]{->}(-4,0)(-3.2,0)
\pscircle*(-4,0){0.2}
\psbezier[linewidth=2pt,linecolor=blue](-3,0)(-1.5,1.5)(-1.5,-1.5)(-3,0)
\pscircle[linewidth=1pt, fillstyle=solid, fillcolor=white](-3,0){0.2}
}
\rput(0,-0.8){
\rput(-5.5,0){\textcolor{black}{$+$}}
\psbezier[linewidth=2pt,linecolor=blue](-5,0)(-4.5,-.6)(-4.5,-.4)(-4,0)
\pcline[linewidth=2pt,linecolor=green]{->}(-5,0)(-4.2,0)
\pscircle[linewidth=1pt, fillstyle=solid, fillcolor=white](-5,0){0.2}
\rput(-5,0){\makebox(0,0){\tcg{$\bullet$}}}
\pcline[linewidth=2pt,linecolor=green]{->}(-4,0)(-3.2,0)
\pscircle[linewidth=1pt, fillstyle=solid, fillcolor=white](-4,0){0.2}
\psbezier[linewidth=2pt,linecolor=red,linestyle=dashed]{->}(-2.8,0.2)(-1.5,1.5)(-1.5,-1.5)(-2.8,-0.2)
\pscircle*(-3,0){0.2}
}
\rput(0,0.8){
\rput(-1,0){\textcolor{black}{$+$}}
\psbezier[linewidth=2pt,linecolor=red,linestyle=dashed]{->}(0,0)(1,.8)(1.2,1)(1.8,0.2)
\psbezier[linewidth=2pt,linecolor=blue](1,0)(1.4,-.5)(1.6,-.5)(2,0)
\pcline[linewidth=2pt,linecolor=green]{->}(0,0)(.8,0)
\pscircle[linewidth=1pt, fillstyle=solid, fillcolor=white](0,0){0.2}
\rput(0,0){\makebox(0,0){\tcg{$\bullet$}}}
\pcline[linewidth=2pt,linecolor=green]{->}(1,0)(1.8,0)
\pscircle[linewidth=1pt, fillstyle=solid, fillcolor=white](1,0){0.2}
\pscircle*(2,0){0.2}
}
\rput(0,-0.8){
\rput(-1,0){\textcolor{black}{$+$}}
\psbezier[linewidth=2pt,linecolor=blue](0,0)(1,.8)(1.2,1)(2,0)
\psbezier[linewidth=2pt,linecolor=red,linestyle=dashed]{->}(1,0)(1.4,-.5)(1.6,-.5)(1.8,-0.2)
\pcline[linewidth=2pt,linecolor=green]{->}(0,0)(.8,0)
\pscircle[linewidth=1pt, fillstyle=solid, fillcolor=white](0,0){0.2}
\rput(0,0){\makebox(0,0){\tcg{$\bullet$}}}
\pcline[linewidth=2pt,linecolor=green]{->}(1,0)(1.8,0)
\pscircle[linewidth=1pt, fillstyle=solid, fillcolor=white](1,0){0.2}
\pscircle*(2,0){0.2}
\psbezier[linewidth=5pt,linecolor=white](0,-0.7)(1,.3)(1,.3)(2,0.7)
\psbezier[linewidth=5pt,linecolor=white](0.1,0.6)(1,0)(1,0)(1.3,-0.8)
\psbezier[linewidth=3pt,linecolor=black](0,-0.7)(1,.3)(1,.3)(2,0.7)
\psbezier[linewidth=3pt,linecolor=black](0.1,0.6)(1,0)(1,0)(1.3,-0.8)
}
\rput(0,1.3){
\rput(3,0){\textcolor{black}{$+$}}
\pcline[linewidth=2pt,linecolor=green]{->}(4,0)(4.65,0.65)
\pcline[linewidth=2pt,linecolor=green]{->}(4,0)(4.65,-0.65)
\psbezier[linewidth=2pt,linecolor=red,linestyle=dashed]{->}(4.8,0.8)(6.3,2.3)(6.3,-.7)(5,0.6)
\psbezier[linewidth=2pt,linecolor=blue](4.8,-0.8)(6.3,-2.3)(6.3,.7)(4.8,-0.8)
\pscircle[linewidth=1pt, fillstyle=solid, fillcolor=white](4,0){0.2}
\rput(4,0){\makebox(0,0){\tcg{$\bullet$}}}
\pscircle[linewidth=1pt, fillstyle=solid, fillcolor=white](4.8,-0.8){0.2}
\pscircle*(4.8,0.8){0.2}
}
\rput(0,-0.8){
\rput(3,0){\textcolor{black}{$+$}}
\psbezier[linewidth=2pt,linecolor=blue](4,0)(4.5,-.6)(4.5,-.4)(5,0)
\pcline[linewidth=2pt,linecolor=green]{->}(4,0)(4.8,0)
\psbezier[linewidth=2pt,linecolor=red,linestyle=dashed]{->}(4.8,0)(6.3,1.5)(6.3,-1.5)(5,-0.2)
\pscircle[linewidth=1pt, fillstyle=solid, fillcolor=white](4,0){0.2}
\rput(4,0){\makebox(0,0){\tcg{$\bullet$}}}
\pscircle[linewidth=1pt, fillstyle=solid, fillcolor=white](5,0){0.2}
}
\end{pspicture}
}
$$
where we explicitly crossed out the diagram that is forbidden by the selection rules. So, the actual answer comprises five
admitted diagrams.

\subsection{Calculating ${\cal F}_{0,2}$}\label{ss:F02}

It was shown in \cite{ChEy-b} that the diagrammatic representation formally applied to ${\cal F}_{0,2}$ just
gives zero.

We can however make a guess for the actual ${\cal F}_{0,2}$. It was proposed in~\cite{ChEy-b} to interpret it
as the Polyakov's gravitational anomaly term $\iint R\frac1{\Delta} R$, where $R$ is the curvature (of
two-dimensional metric)
and $1/\Delta$ is the Green's function for
the Laplace operator, which in our case is the logarithm of the Prime form. The curvature is expressed through
the function $y$ as $R\sim y'/y$. That is, we have two natural candidates for ${\cal F}_{0,2}$:
$$
{\cal F}_{0,2}\sim \iint dq\,dp\frac{y'(q)}{y(q)}\log E(p,q)\frac{y'(p)}{y(p)},
$$
where $E$ is the Prime form, or
$$
{\cal F}_{0,2}\sim \iint dq\,dp\log y(q) B(p,q)\log y(p),
$$
but neither of these expressions is well defined. The first one develops the logarithmic cut at $p=q$ and cannot be
written in a contour-independent way; moreover, both these expressions are divergent when integrating along the
support. We therefore must find another representation imitating this term. A proper choice turns out to be the one
in which we integrate by part only once therefore producing the expression of the form
\be
\oint_{{\cal C}_{\cal D}^{(q)}}\frac{dq}{2\pi i}
\frac{y'(q)}{y(q)}\int_{\cal D}dE_{q,\bar q}(p)\log y(p) dp,
\label{F02:3}
\ee
where the second integral is taken exactly along the eigenvalue support $\cal D$.

Variation of the logarithmic term in (\ref{F02:3}) can be presented already in the form
of the contour integral. Referring the reader for details to \cite{ChEy-b}, we present
here only the result.

Introducing $M_{\alpha}^{(1)}\equiv \left.Q(p)\right|_{p=\mu_\alpha}$ to be the first {\em moments}
of the total potential $V$ and $\Delta(\mu)$ to be the Vandermonde determinant of the branch points $\mu_\alpha$,
we have
\be
\label{F02:6}
{\cal F}_{0,2}=
-\oint_{{\cal C}_{\cal D}^{(q)}}\frac{dq}{2\pi i}\frac{y'(q)}{y(q)}\int_{D}dE_{q,\bar q}(p)\log y(p) dp
-\frac13\log\left(\prod_{\alpha=1}^{2n}M_{\alpha}^{(1)}\Delta(\mu)\right),
\ee
and we indeed have {\em quantum correction} term (the second term in (\ref{F02:6})).

\subsection*{Acknowledgments}
The author thanks Bertrand Eynard for collaboration, Sara Pasquetti and Marcos Mari{\~n}o for the useful discussion,
Gabriel Cardoso for the hospitality during my visit to the Program in Matrix Models in Instituto Superior T\'ecnico, Lisbon,
Portugal, and Eyjafjallajokull for ensuring the possibility of this long discussion.

The work is supported by the Russian Foundation for Basic Research (Grants 10-02-01315$\_$a,
10-01-92104-JF$\_$a and 09-01-12150-ofi$\_$m), by the Grant for Supporting Leading Scientific Schools NSh-795.2008.1,
and by the Program Mathematical Methods for Nonlinear Dynamics.


\begin{thebibliography}{100}

\bibitem{AGT} L.~F.~Alday, D.~Gaiotto, and Y.~Tachikawa, {\it Liouville correlation function from four-dimensional
gauge theories}, {\sl Lett. Math. Phys.} {\bf 91} (2010) 167-197; arXiv:0906.3219 [hep-th].

\bibitem{BEMN}
G. Borot, B. Eynard, S. Majumdar, and C. Nadal, {\it Large deviations of the maximal eigenvalue of random matrices},
arXiv:1009:1945 [math-ph].

\bibitem{Ch1} L. Chekhov,
{\it Genus one corrections to multi-cut matrix model solutions},
{\sl Theor. Math. Phys. } {\bf 141} (2004) 1640--1653, hep-th/0401089.


\bibitem{ChEy}
L. Chekhov, B. Eynard, {\it  Hermitean matrix model free energy:
Feynman graph technique for all genera}, {\sl JHEP} {\bf 0603}:014
(2006).

\bibitem{ChEy-b}
L. Chekhov and B. Eynard, {\it Matrix eigenvalue model: Feynman
graph technique for all genera}, {\sl JHEP} 12(2006)026.


\bibitem{DV09} R. Dijkgraaf and C. Vafa, {\it Toda theories, matrix models, topological strings, and $N=2$ gauge systems},
arXiv:0909.2453 [hep-th].

\bibitem{Eguchi} T. Eguchi and K. Maruyoshi, {\it Penner-type matrix model and Seiberg--Witten theory},
{\sl JHEP} {\bf 1002}(2010)022; arXiv:0911.4797 [hep-th].

\bibitem{Ey1}
B.Eynard,
{\it Topological expansion for the 1-Hermitian matrix model correlation functions}, {\sl JHEP} {\bf 0411}(2004)031,
hep-th/0407261.




\bibitem{MMM} A.~Mironov and A.~Morozov, {\it The power of Nekrasov functions}, Phys. Lett. B{\bf680}, 188 (2009) [arXiv:0908.2190 [hep-th]].\\
            A.~Marshakov, A.~Mironov and A.~Morozov, {\it On non-conformal limit of the AGT relations} arXiv:0909.2052 [hep-th].\\
            A.~Marshakov, A.~Mironov and A.~Morozov, {\it Zamolodchikov asymptotic formula and instanton expansion in $N=2$ SUSY $N_f=2N_c$ QCD},
            arXiv:0909.3338 [hep-th].\\


\end{thebibliography}
\end{document}